\documentclass[
    acmtog, %
    authorversion=true,
    screen=true,
    authordraft=false,
    review=false,
    timestamp=false,
    anonymous=false, 
    nonacm=true]{acmart}

\AtBeginDocument{%
	\providecommand\BibTeX{{%
			\normalfont B\kern-0.5em{\scshape i\kern-0.25em b}\kern-0.8em\TeX}}}

\setcopyright{rightsretained}
\copyrightyear{2025}
\begin{document}
	
\title{FlatCAD: Fast Curvature Regularization of Neural SDFs for CAD Models}
	
\author{Haotian Yin}
\affiliation{
    \institution{New Jersey Institute of Technology}
    \country{United States}
}
\authornote{Authors contributed equally to this work.}

\email{hy9@njit.edu}

\author{Aleksander Plocharski}
\affiliation{
    \institution{Warsaw University of Technology}                      
    \country{Poland}  
}
\affiliation{    
    \institution{IDEAS NCBR}    
    \country{Poland}             
}

\authornotemark[1]        %
\email{aleksander.plocharski@pw.edu.pl}

\author{Michał Jan Włodarczyk}
\affiliation{
    \institution{Warsaw University of Technology}  
    \country{Poland}             
}
\author{Mikołaj Kida}
\affiliation{
    \institution{Warsaw University of Technology}  
    \country{Poland}             
}

\author{Przemyslaw Musialski}
\affiliation{
    \institution{New Jersey Institute of Technology}  
    \country{United States}             
}
\affiliation{
    \institution{IDEAS Research Institute}  
    \country{Poland}             
}
\email{przem@njit.edu}

\begin{abstract}
Neural signed-distance fields (SDFs) are a versatile backbone for neural geometry representation, but enforcing CAD-style developability usually requires Gaussian-curvature penalties with full Hessian evaluation and second-order differentiation, which are costly in memory and time. We introduce an off-diagonal Weingarten loss that regularizes only the mixed shape operator term that represents the gap between principal curvatures and flattens the surface. We present two variants: a finite-difference version using six SDF evaluations plus one gradient, and an auto-diff version using a single Hessian-vector product. Both converge to the exact mixed term and preserve the intended geometric properties without assembling the full Hessian. On the ABC benchmarks the losses match or exceed Hessian-based baselines while cutting GPU memory and training time by roughly a factor of two. The method is drop-in and framework-agnostic, enabling scalable curvature-aware SDF learning for engineering-grade shape reconstruction. Our code is available at \url{https://flatcad.github.io/}. 
\end{abstract}



\begin{teaserfigure}
 \centering
    \includegraphics[width=0.99\linewidth]{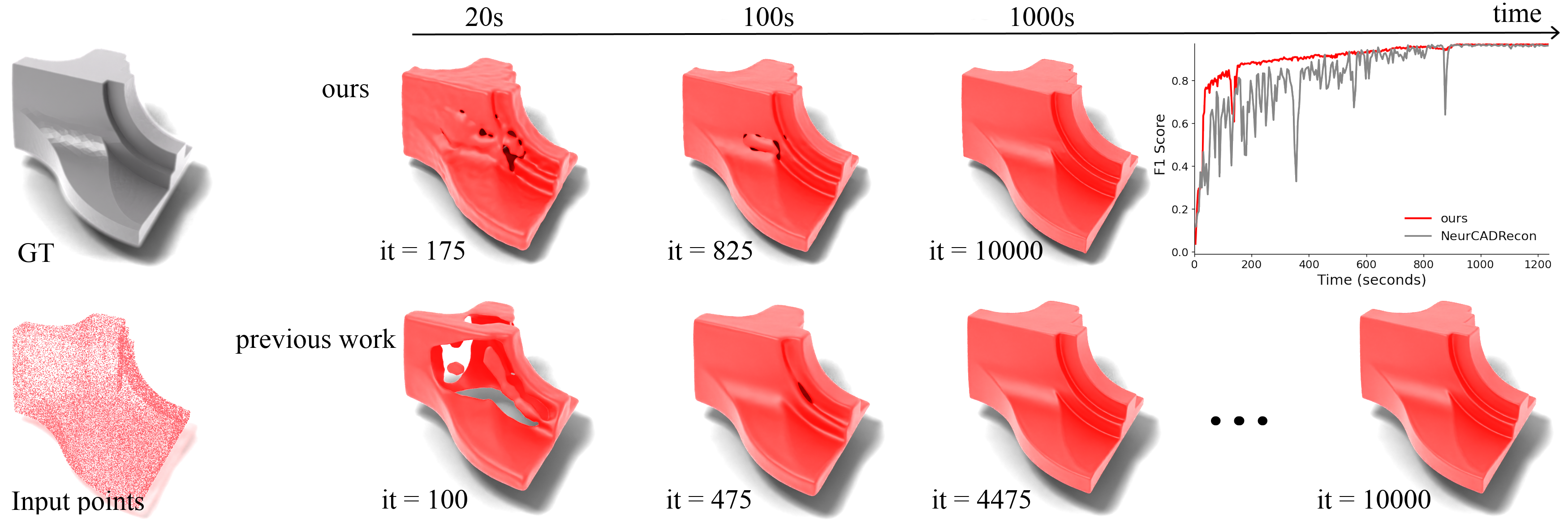}
    \caption{Comparison of our Weingarten regularizer (top row) and the full-Hessian-based method of previous work~\cite{Dong2024NeurCADReconNR} (bottom row) on a CAD part reconstruction: both achieve similar-quality surfaces, but our method converges in roughly half the time and uses about half the memory.}
    \label{fig:teaser}
\end{teaserfigure}

\maketitle

\section{Introduction}\label{sec:intro}

\newcommand{\SIREN}{{SIREN}~}

Neural implicit representations of geometry, specifically \textit{signed-distance fields} (SDFs) encoded by multilayer perceptrons have become a novel and powerful representation in 3D vision and graphics. Early work \cite{Park2019DeepSDFLC,Mescheder2018OccupancyNL} showed that a modest MLP could store a watertight shape and support arbitrary resolution through continuous evaluation and differentiation.  
That promise was broadened by \SIREN \cite{Sitzmann2020ImplicitNR}, whose sinusoidal activations capture fine geometric detail and high-frequency signals, and positional encodings that brought ReLU-based networks to high detail accuracy~\cite{Tancik2020FourierFL} and essentially real-time performance~\cite{Mller2022InstantNG}. 

However, for accurate geometric representations, a remaining challenge was {geometric faithfulness}: vanilla SDF fits can satisfy point-wise losses while remaining wildly inconsistent in surface detail or curvature. Lipman et al. addressed this with the {Implicit Geometric Regularizer} (IGR) \cite{Gropp2020ImplicitGR}, adding an Eikonal term $\lVert \nabla f \rVert \approx 1$ that enforces unit-length gradients, followed by second-order methods like {DiGS} \cite{BenShabat2021DiGSD}, which stabilizes training using the divergence, and {Neural-Singular-Hessian} \cite{Wang2023NeuralSingularHessianIN}, which pushes the network’s Hessian toward low rank, which discourages spurious curvature and enforces piece-wise smoothness that matches better with classical differential geometry.

That thread naturally intersects with computer-aided design (CAD). Mechanical parts are dominated by plates, cylinders, cones, and developable blends whose {Gaussian curvature is zero almost everywhere}, except for sharp feature curves. Encoding these properties in an SDF prior promises more faithful reconstructions from sparse or noisy scans and cleaner patch decomposition for downstream parametric recovery. {NeurCADRecon}~\cite{Dong2024NeurCADReconNR} recently embodied that idea: it augments a \SIREN network  with a ``developability'' loss that minimizes the magnitude of Gaussian curvature $|K|$ over a thin shell around the surface, thereby encouraging each learned patch to flatten out unless data dictates otherwise.

Unfortunately, such elegance comes at a price. The Gaussian curvature of an implicit surface depends on {all nine entries of the Hessian of the network}. In automatic differentiation frameworks this means six Jacobian-vector products and thus seven backward passes {per sample}, plus the retention of second-order computational graphs. Memory footprints balloon, forcing small batches; iteration times rise, limiting practical resolutions. 

To address these limitations we propose a novel curvature regularization loss. In particular, \textbf{our contributions} are the following: 
\begin{itemize}    

  \item \textbf{Curvature gap regularization.} We regularize only the off-diagonal Weingarten shape operator term, letting each principal curvature follow the data while flattening hyperbolic and parabolic patches and rounding elliptic ones uniformly. In other words, we minimize the ``curvature-gap'' which is the difference between the principal curvatures and we denote the resulting regularizer the \textit{off-diagonal Weingarten loss} (ODW-loss). It is described in detail in Section~\ref{sec:method}. 

  \item \textbf{Two implementations.}
  We propose two variants for computing the regularizer: 
  (i) Finite‐differences, which uses forward and backward offsets (six additional SDF queries plus one gradient evaluation) and introduces no second-order graphs; 
  (ii) Auto-differentiation, which employs a single Hessian-vector product obtained with two reverse sweeps, thereby avoiding full-Hessian assembly and running faster in practice.
  Both are described in detail in Section~\ref{sec:pathways}.

  \item \textbf{Efficiency without loss of accuracy.} On the ABC CAD benchmark our proxies match or slightly outperform Hessian-based baselines while roughly halving GPU memory use and training time. The details are presented  in Section~\ref{sec:experiments}. 
\end{itemize}

Because the method is drop‐in and framework‐agnostic, it makes curvature‐aware neural SDFs practical for large‐scale, engineering‐grade shape reconstruction. Our implementation is available at \url{https://flatcad.github.io/}.

\section{Related Work}\label{sec:related}

\paragraph{Implicit Surface Representations and Reconstruction}

Implicit representations have emerged as a powerful alternative to explicit surface models due to their continuous and differentiable nature. Reconstructing surfaces from point clouds has long been a fundamental task in computer graphics, and recent years have witnessed a surge in learning-based implicit approaches.

Early implicit methods compute the signed distance to the tangent plane of the closest surface point~\cite{Hoppe1992SurfaceRF}. Subsequently, radial basis function (RBF) methods were introduced to fit the zero level-set of the signed distance function, enabling smooth surface reconstruction~\cite{Carr2001ReconstructionAR, Huang2019VariationalIP, Li2016SparseRS}. Another family of methods formulates reconstruction as solving partial differential equations, such as Poisson Surface Reconstruction (PSR)~\cite{Kazhdan2020PoissonSR,Kazhdan2013ScreenedPS,Kazhdan2006PoissonSR,Hou2022IterativePS}, which computes an occupancy field by solving a Poisson equation. Parametric Gauss Reconstruction (PGR)~\cite{Lin2022SurfaceRF} improves normal consistency by leveraging the Gauss formula from potential theory. These methods build surfaces analytically by imposing strong geometric constraints.

In contrast, neural implicit methods learn to map 3D coordinates to continuous scalar fields, representing the surface as the zero level-set of a neural network. DeepSDF~\cite{Park2019DeepSDFLC} pioneered the use of an auto-decoder that optimizes a latent code per shape using MAP estimation. Supervised learning-based approaches extend this idea by training on datasets with precomputed ground-truth field values, using structured representations such as voxel grids~\cite{Chabra2020DeepLS,Jiang2020LocalIG,Peng2020ConvolutionalON}, k-nearest neighbor graphs~\cite{Boulch2022POCOPC,Erler2020Points2SurfLI}, and octrees~\cite{Huang2022ANG,Tang2021OctFieldHI,Wang2022DualOG}. Some methods target open surfaces or unsigned distance fields (UDFs)~\cite{Chibane2020ImplicitFI,Ye2022GIFSNI}. While these methods can reconstruct detailed shapes, they often encode a fixed representation for a family of similar shapes and thus generalize poorly to unseen geometries.

To address generalization and robustness, recent methods directly optimize a neural field from raw point clouds in a self-supervised fashion. These methods typically rely on regularization terms that encode geometric priors. For example, IGR\cite{Gropp2020ImplicitGR} introduces the Eikonal regularization to encourage unit-norm gradients, a defining property of valid distance fields. SAL and SALD\cite{Atzmon2019SALSA} adopt sign-agnostic learning to regress signed distances from unsigned data. \SIREN~\cite{Sitzmann2020ImplicitNR} uses periodic activation functions to preserve high-frequency details. 

To further enhance geometric fidelity, higher-order constraints have also been explored. The Hessian matrix of a neural field encodes its second-order derivatives, making it a natural vehicle for expressing differential properties such as surface curvature. DiGS~\cite{BenShabat2021DiGSD} penalizes the divergence of the gradient field as a soft constraint, encouraging smooth transitions across the surface. Neural-Singular-Hessian~\cite{Wang2023NeuralSingularHessianIN} directly regularizes the Hessian by minimizing the smallest singular value to make it rank-deficient near the surface, promoting developability and reducing spurious oscillations. These second-order constraints have proven effective in restoring fine-grained geometric structures. 
Concurrently, other curvature-guided regularizers have been explored for surface INRs \cite{Sang2025_Normals,Novello_2023_ICCV}, and rank-minimization schemes have been proposed to approximate developability \cite{Selvaraju2024_Developa}.

\paragraph{Surface Reconstruction of CAD Models}
In the context of CAD model reconstruction, several approaches have focused on fitting predefined geometric primitives. ~\cite{Kania2020UCSGNetU} and ~\cite{Sharma2017CSGNetNS} adopt primitive fitting pipelines, while~\cite{Li2018SupervisedFO} employ supervised learning to detect primitive types and fit patches such as planes, cylinders, and spheres. ~\cite{Sharma2020ParSeNetAP} extend this idea to B-spline patches using differentiable segmentation, and~\cite{Uy2021Point2CylRE} segment point clouds into extrusion cylinders assembled via Boolean operations. However, these methods often require voxelized and oriented inputs, which makes them less practical for handling complex CAD assemblies that are scanned as raw, unoriented point clouds.

More recently, implicit neural representations have also been applied to CAD reconstruction under a self-supervised setting. ~\cite{Dong2024NeurCADReconNR} proposes a novel formulation that integrates curvature-based regularization into the learning process. Specifically, it introduces a Gaussian curvature loss that encourages zero curvature across the reconstructed surface, based on the assumption that CAD models are composed of piecewise-developable patches. This curvature regularization is expressed through a differentiable formulation involving the Hessian of the neural SDF.

\begin{figure*}[t]
    \centering
    \includegraphics[width=0.99\linewidth]{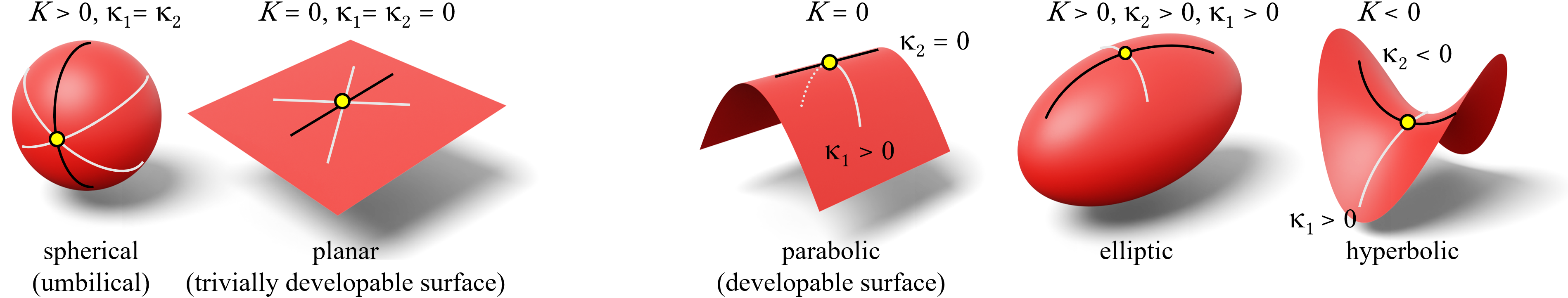}
    \caption{Principal curvatures $\kappa_1$ and $\kappa_2$ are the eigenvalues of the Weingarten map $S$. Their product gives the Gaussian curvature $K=\kappa_1 \kappa_2 = \det S$. Because $S$ is self-adjoint with respect to the first fundamental form, its eigenvectors (the principal directions) are orthogonal in the surface metric whenever the eigenvalues are distinct. At an umbilic point (e.g. on a sphere) the entire two-dimensional tangent plane is the eigenspace, so no unique pair of principal directions exists. For a plane the second fundamental form vanishes, i.e., $S=0$.}
    \label{fig:curvature}
\end{figure*}

\section{Method}\label{sec:method}

Our goal is to learn a neural signed-distance field $f$ from an unoriented point cloud such that its zero-set reproduces the target CAD surface. We keep the standard supervision recipe used by implicit reconstruction methods and add a minimal curvature regularizer that removes local warp, flattening developable and saddle parts and rounding elliptic ones--without ever building a full Hessian.

First, we inherit standard supervision terms commonly used in implicit learning: a Dirichlet condition on the zero-level set, as proposed by Atzmon et al.~\cite{Atzmon2020SALSA}, and the Eikonal constraint~\cite{Gropp2020ImplicitGR}, which enforces unit gradient norms in the vicinity of the surface. These terms ensure fidelity to the input data and preserve geometric consistency.

Second, and most importantly, we introduce a novel curvature regularizer that suppresses the difference between the principal curvatures  of the learned surface. This regularization is particularly effective for CAD surfaces, as it forces parabolic and hyperbolic surface parts to be planar and elliptic areas to be umbilical (cf. Figure~\ref{fig:curvature}). In combination with the surface reconstruction (Dirichlet) and unit gradient field (Eikonal) losses, it promotes flatness where the target is hyperbolic and parabolic, and spherical uniformity where it is doubly curved. 

Unlike previous methods that construct the full $ 3\times 3$ Hessian~\cite{Wang2023NeuralSingularHessianIN, Dong2024NeurCADReconNR}, our formulation needs only a single Hessian–vector product per sample, eliminating explicit Hessian assembly and speeding up training. Additionally, we also propose a first-order finite-difference approximation, which avoids Hessian computation entirely, enabling significantly faster training while retaining the geometric benefits of curvature regularization.

\subsection{Signed-Distance Fields and Surface Geometry}

We represent a surface $\mathcal{M}\subset\mathbb{R}^{3}$ implicitly as the zero-level set of a signed-distance field (SDF)
\[
f:\mathbb{R}^{3}\!\longrightarrow\!\mathbb{R},
\qquad
\mathcal{M}=\bigl\{\mathbf{x}\mid f(\mathbf{x})=0\bigr\}.
\]
Because a true SDF satisfies $\lVert\nabla f\rVert=1$ in a neighborhood of $\mathcal{M}$, its \emph{normal direction is locally linear}:
\[
\partial_{nn}f
  =\mathbf{n}^{\!\top}H_f\,\mathbf{n}=0,
\qquad
\mathbf{n}=\frac{\nabla f}{\lVert\nabla f\rVert},
\]
where $H_f=[\partial_{x_i}\partial_{x_j}f]$ is the Hessian of the SDF.  
The shape-operator (Weingarten map) is obtained by projecting $H_f$ into any orthonormal tangent frame $(\mathbf{u},\mathbf{v})$:
\[
S=
\begin{pmatrix}
\mathbf{u}^{\!\top}H_f\,\mathbf{u} & \mathbf{u}^{\!\top}H_f\,\mathbf{v}\\
\mathbf{v}^{\!\top}H_f\,\mathbf{u} & \mathbf{v}^{\!\top}H_f\,\mathbf{v}
\end{pmatrix}. 
\]
That operator (Weingarten map) takes any  tangent vector and returns how fast the surface normal rotates in that direction, i.e., the local bending.  
Its eigenvalues are the \emph{principal curvatures} $\kappa_1,\kappa_2$; their product gives the Gaussian curvature $K=\kappa_1\kappa_2=\det S$. Together, these values distinguish spherical, planar, parabolic, elliptic, and hyperbolic regimes as illustrated in Fig.~\ref{fig:curvature}.

\paragraph{Rotation of $S$ and the Mixed Entry}
In the principal frame the shape operator is
\[
S_0=\begin{pmatrix}\kappa_1&0\\[2pt]0&\kappa_2\end{pmatrix}.
\]
\noindent
We can rotate a tangent basis by an angle $\theta$ through 
\[
R(\theta)=
\begin{pmatrix}
\cos\theta&-\sin\theta\\
\sin\theta&\phantom{-}\cos\theta
\end{pmatrix},
\qquad R^{\!\top}R=I .
\]
\noindent
The operator in the rotated frame is the orthogonal transform
\[
S(\theta)=R^{\!\top}S_0R
         =\begin{pmatrix}
            \kappa_1\cos^{2}\theta+\kappa_2\sin^{2}\theta &
            (\kappa_2-\kappa_1)\sin\theta\cos\theta\\[4pt]
            (\kappa_2-\kappa_1)\sin\theta\cos\theta &
            \kappa_1\sin^{2}\theta+\kappa_2\cos^{2}\theta
           \end{pmatrix}.
\]
\noindent
The off-diagonal component is
\begin{equation}
S_{12}(\theta)=
(\kappa_2-\kappa_1)\sin\theta\cos\theta
 =\tfrac12(\kappa_2-\kappa_1)\sin 2\theta .
 \label{eq:s12}
\end{equation}
Thus $S_{12}$ is a \emph{warp} measure: it vanishes in a principal frame and is proportional to the curvature difference in any other frame. 

\subsection{Maximum Likelihood Estimation}
During training we penalize $S_{12}^2$  (or $|S_{12}|$, see below), in particular, for each point we 
sample a fresh random frame angle $\theta$ each iteration (which in practice is equivalent to sampling a fresh random vector $\mathbf{u}$ orthogonal to $\mathbf{n}$, and completing the frame with $\mathbf{v}$ using the cross product).

Taking the expectation of $S_{12}^{2}$ over $\theta\in[0,2\pi)$ yields
\[
\mathbb{E}_{\theta}\bigl[S_{12}^2(\theta)\bigr]
   =\frac{(\kappa_2-\kappa_1)^2}{4}\,
     \mathbb{E}_{\theta}\bigl[\sin^22\theta\bigr].
\]
Because
\[
\mathbb{E}_{\theta}\bigl[\sin^22\theta\bigr]
   =\frac{1}{2\pi}\int_0^{2\pi}\sin^22\theta\,d\theta
   =\frac12,
\]
we obtain
\begin{equation}
\mathbb{E}_{\theta}[S_{12}^{2}]
      =\frac{1}{8}(\kappa_2-\kappa_1)^2 .
\label{eq:mean_s12}
\end{equation}

Equation~\eqref{eq:mean_s12} vanishes \emph{iff} $\kappa_1=\kappa_2$.  
Hence gradient descent on $S_{12}^2$ systematically \emph{suppresses the curvature gap} without directly pushing either curvature to zero.

\noindent
Because the frame angle is resampled independently as $\theta\sim\mathcal U[0,2\pi)$, the off-diagonal Weingarten entry $S_{12}$
is a zero-mean random variable.
The mini-batch average of $S_{12}^{2}$ (or $|S_{12}|$) is thus an unbiased Monte-Carlo estimate of its expectation over $\theta$.
If the measurement noise on $S_{12}(\theta)$ is assumed Gaussian with variance $\sigma^{2}$, the negative log-likelihood reduces---up to a constant---to 
\[
\tfrac1{2\sigma^{2}}\sum S_{12}(\theta)^{2} \;. 
\]
Minimizing the squared off-diagonal entry is therefore the ma\-xi\-mum-like\-li\-hood estimator for the zero-warp hypothesis. Consequently, the $L^2$ formulation possesses a clear maximum-likelihood interpretation under the common Gaussian-noise assumption.

\paragraph{Expectation of the absolute value}
Adopting a Laplace noise model instead yields the absolute-value, with the same Monte-Carlo interpretation. With \(\theta\sim\mathrm{Uniform}[0,2\pi)\),
\[
\begin{aligned}
\mathbb{E}_{\theta}\bigl[\lvert S_{12}(\theta)\rvert\bigr]
   &\,=\,\lvert\kappa_2-\kappa_1\rvert\;
          \mathbb{E}_{\theta}\bigl[\lvert\sin\theta\cos\theta\rvert\bigr] \\[6pt]
   &\,=\,\lvert\kappa_2-\kappa_1\rvert\;
          \frac{1}{2\pi}\int_{0}^{2\pi}\!\lvert\sin\theta\cos\theta\rvert\,d\theta .
\end{aligned}
\]
Evaluating the integral:
\[
\int_{0}^{2\pi}\!\lvert\sin\theta\cos\theta\rvert\,d\theta
  =\frac12\int_{0}^{2\pi}\!\lvert\sin 2\theta\rvert\,d\theta
  =\frac12\!\times\!4
  =2 
\]
yields 
\begin{equation}
    \mathbb{E}_{\theta}\bigl[\lvert S_{12}\rvert\bigr]
      =\frac{1}{\pi}\lvert\kappa_2-\kappa_1\rvert .
\end{equation}

While the squared Weingarten loss has elegant Gaussian-noise and variance interpretations, in a \SIREN setting the practical stability, edge preservation, and faster optimization of the absolute-value off-diagonal Weingarten loss generally outweigh the theoretical neatness of $L^2$.

\SIREN network's high-frequency sine layers can generate very large local gradients; employing the absolute ($L^1$) loss caps each sample’s gradient at $\pm 1$, preventing exploding updates while preserving sharp detail, and thus trains more stably than the squared ($L^2$) version.

\subsection{Hessian Deficiency}
Evaluating $S_{12}$ directly on the zero level-set is numerically unstable: the Hessian is rank-deficient because $\partial_{nn}f = 0$. 
In fact, since the Eikonal condition enforces $\|\nabla f\|=1$, differentiating this constraint along the normal direction yields
\[
    \partial_{n}\!\left(\tfrac{1}{2}\|\nabla f\|^{2}\right) 
      \;=\; (H_f \mathbf{n}) \cdot \nabla f 
      \;=\; \mathbf{n}^{\top} H_f \mathbf{n} 
      \;=\; 0,
\]
which means that one eigenvalue of $H_f$ must vanish, making the Hessian singular on the surface. 
This property of signed distance fields is also emphasized by Wang et al.~\cite{Wang2023NeuralSingularHessianIN}, who use it as the basis for their Neural-Singular-Hessian regularizer.

To address the numerical instability, we follow the sampling strategy of Dong et al.~\cite{Dong2024NeurCADReconNR}. 
At each iteration, we sample a static shell $\Omega$ by drawing one point around every input point $p$ from a 3D Gaussian whose standard deviation equals the distance to its 50-th nearest neighbor, yielding a fixed batch of 15k near-surface points. 
We then evaluate $S_{12}$ on $\Omega$, which provides stable signals while leaving the standard data and Eikonal terms unchanged.

The off-diagonal Weingarten loss is hence a minimal, orientation-free condition that suppresses the curvature difference $(\kappa_2-\kappa_1)^2$, thereby isolating the spurious ``warp'' defect and flattening hyperbolic and parabolic areas. 

Because the loss depends only linearly on this mixed second derivative, rather than on quadratic combinations of Hessian entries, it yields well-conditioned, low-variance gradients that integrate seamlessly with standard SDF and Eikonal losses, mitigating the exploding sensitivities observed with many quadratic-or-higher curvature regularizers.

\subsection{Loss Function Components}
\label{subsec:loss}

Our training objective combines four terms that constrain the signed‐distance field both on and off the surface.  
The total loss is
\begin{equation}
\mathcal{L}_{\text{total}}
  = \lambda_{\text{DM}}\mathcal{L}_{\text{DM}}
  + \lambda_{\text{DNM}}\mathcal{L}_{\text{DNM}}
  + \lambda_{\text{EIK}}\mathcal{L}_{\text{EIK}}
  + \lambda_{\text{ODW}}\mathcal{L}_{\text{ODW}},
\end{equation}
with scalar weights $\lambda_{\text{DM}}$, $\lambda_{\text{DNM}}$, $\lambda_{\text{EIK}}$, and $\lambda_{\text{ODW}}$ balancing the contributions of the three regularizers against the data term.

\paragraph{Manifold Loss}
For the $N$ input points $\{\mathbf{x}_i\}_{i=1}^{N}$ sampled on or very near the target surface, we impose a Dirichlet condition that pulls their SDF values to~\(0\):
\begin{equation}
\mathcal{L}_{\text{DM}}
  = \frac{1}{N}\sum_{i=1}^{N}\bigl|f(\mathbf{x}_i)\bigr|.
\end{equation}
This term anchors the learned zero‐level set to the point cloud and provides first‐order geometric fidelity.

\paragraph{Non‐Manifold Loss}
For $M$ free‐space samples $\{\mathbf{y}_j\}_{j=1}^{M}$ we adopt the sign‐agnostic formulation of Atzmon et al.~\cite{Atzmon2020SALSA}:
\begin{equation}
\mathcal{L}_{\text{DNM}}
  = \frac{1}{M}\sum_{j=1}^{M}\exp\!\bigl(-\alpha\,|f(\mathbf{y}_j)|\bigr),
  \qquad\alpha=100.
\end{equation}
The exponential quickly attenuates the penalty as \(|f|\) grows, suppressing spurious oscillations close to the surface while allowing the SDF to expand smoothly in regions devoid of geometric evidence, thereby stabilizing training and preventing distant artifacts. 

\paragraph{Eikonal Loss}
To preserve the metric property of a true distance field we enforce the Eikonal equation on $K$ uniformly sampled points $\{\mathbf{z}_k\}_{k=1}^{K}$:
\begin{equation}
\mathcal{L}_{\text{EIK}}
  = \frac{1}{K}\sum_{k=1}^{K}\Bigl(\,\lVert\nabla f(\mathbf{z}_k)\rVert_2-1\Bigr)^{2}.
\end{equation}
This global term regularizes gradient norms, stabilizing optimization and improving extrapolation away from the data~\cite{Gropp2020ImplicitGR}.

\paragraph{Off-Diagonal Weingarten Loss (ours)}
Let $S_{12}(\theta)$ denote the off‐diagonal entry of the $2{\times}2$ shape operator obtained by projecting the Hessian onto a tangent frame rotated by a random angle $\theta\sim\mathcal U[0,2\pi)$.  
For $L$ near‐surface samples $\{\mathbf{p}_\ell\}_{\ell=1}^{L}$ we penalize the \emph{absolute} curvature gap,
\begin{equation}
\mathcal{L}_{\text{ODW}}
  = \frac{1}{L}\sum_{\ell=1}^{L}\bigl|S_{12}(\theta_\ell)\bigr|.
\end{equation}
The mixed second derivative $S_{12}$ is evaluated either with a single Hessian–vector product or via the symmetric finite-difference approximation, avoiding explicit Hessian assembly.

Minimizing $|S_{12}|$ drives the principal curvatures toward equality, flattening parabolic and hyperbolic regions and rounding elliptic regions uniformly.  In all benchmarks the absolute‐value variant yields the most stable optimization and the lowest validation error, so we use it by default in all experiments.

\smallskip
\noindent\textbf{Weights.}  
We keep most parameter weights identical to the settings recommended in {NeurCADRecon}: $\lambda_{\text{DM}}=7000$, $\lambda_{\text{DNM}}=600$, $\lambda_{\text{EIK}}=50$, and \(\lambda_{\text{ODW}}=10\).

\section{Computational Pathways}\label{sec:pathways}

The mixed–curvature term \(S_{12}\) can be evaluated with or without explicit second-order auto-differentiation, giving two practical routes that trade accuracy for running cost. 
The finite-difference minimizes memory use, whereas the auto-diff variant offers higher accuracy. Both achieve the curvature-gap regularization without assembling the full Hessian.  

\begin{figure}[t]
    \centering
    \includegraphics[width=0.99\linewidth]{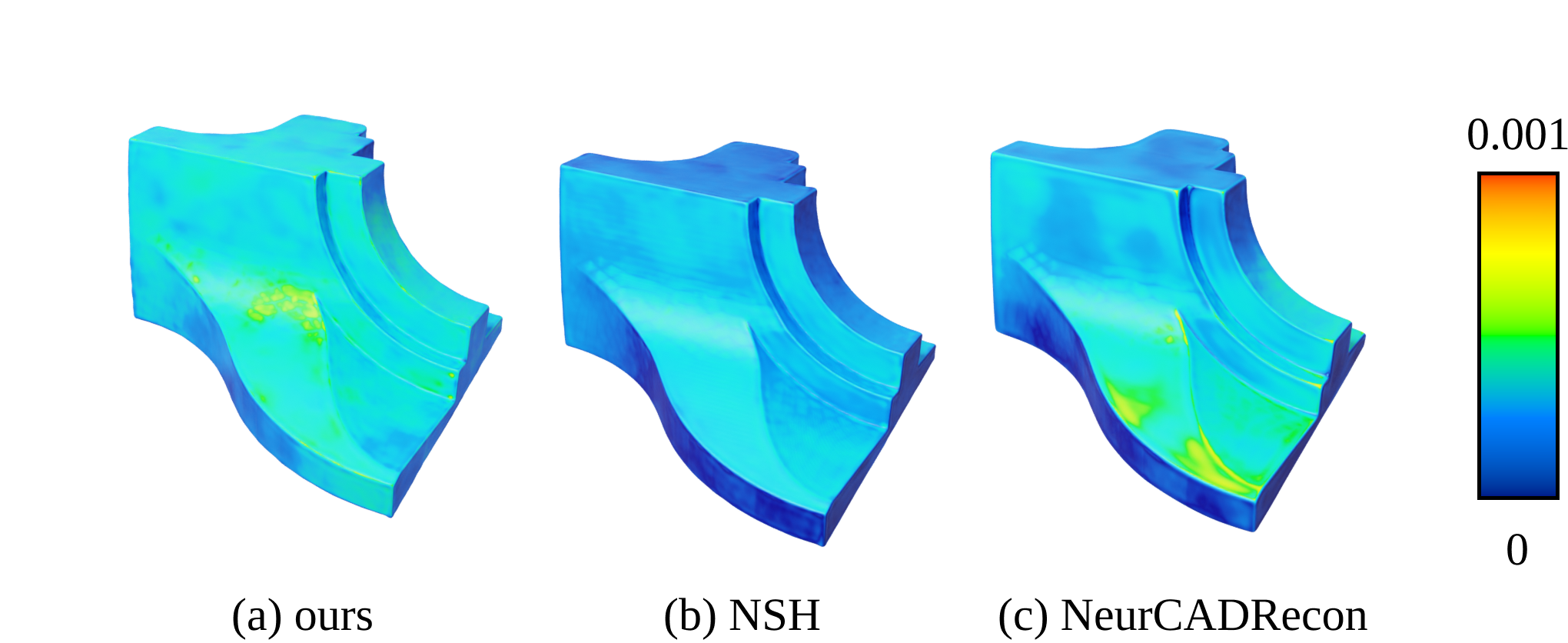}
\caption{Hausdorff–distance heat maps (blue $\rightarrow$ low, yellow $\rightarrow$ high) for the fandisk model reconstructed with ours, NSH, and NeurCADRecon.}
    \label{fig:hausdorff}
\end{figure}

\subsection{Off-Diagonal Weingarten Loss via Taylor Expansion}
\label{subsec:first_order_odw}

When raw throughput or framework simplicity is paramount we approximate the mixed derivative with a symmetric finite-difference stencil.  This stencil evaluates the SDF at forward and backward offsets of magnitude \(h\), averages the two estimates, and therefore incurs a truncation error of order \(h^2\). 
If only the one-sided (forward) stencil is used, 
the truncation error is $\mathcal{O}(h)$ and the estimate is biased. Our default uses the symmetric average which cancels the odd powers and yields $\mathcal{O}(h^2)$ accuracy.

Let \(x_\Omega\) be any point in the near-surface shell \(\Omega\), and let \(\mathbf{u},\mathbf{v}\) form an orthonormal tangent frame at \(x_\Omega\).
For a small step \(h>0\) define the forward offsets
\begin{alignat*}{2}
  f_{00} &= f(x_\Omega), &\quad
  f_{\mathbf{u}}^{+} &= f(x_\Omega + h\,\mathbf{u}),\\[-0.2em]
  f_{\mathbf{v}}^{+} &= f(x_\Omega + h\,\mathbf{v}), &\quad
  f_{\mathbf{uv}}^{+} &= f(x_\Omega + h\,(\mathbf{u}+\mathbf{v})),
\end{alignat*}
and the backward offsets
\begin{alignat*}{2}
  f_{\mathbf{u}}^{-} &= f(x_\Omega - h\,\mathbf{u}), &\quad
  f_{\mathbf{v}}^{-} &= f(x_\Omega - h\,\mathbf{v}),\\[-0.2em]
  f_{\mathbf{uv}}^{-} &= f(x_\Omega - h\,(\mathbf{u}+\mathbf{v})).
\end{alignat*}
Define the one-sided mixed differences and their symmetric average
\begin{equation*}
\begin{aligned}
  D_{\mathbf{uv}}^{(+)}
  &= \frac{f_{\mathbf{uv}}^{+} - f_{\mathbf{u}}^{+} - f_{\mathbf{v}}^{+} + f_{00}}{h^{2}},\\
  D_{\mathbf{uv}}^{(-)}
  &= \frac{f_{\mathbf{uv}}^{-} - f_{\mathbf{u}}^{-} - f_{\mathbf{v}}^{-} + f_{00}}{h^{2}},\\
  D_{\mathbf{uv}}^{(\mathrm{c})}
  &= \tfrac12\!\left(D_{\mathbf{uv}}^{(+)} + D_{\mathbf{uv}}^{(-)}\right).
\end{aligned}
\end{equation*}

\noindent\textbf{Taylor justification (order and bias).}
Assume \(f\in C^3\) near \(x_\Omega\). Using
\[
f(x_\Omega+h\mathbf{a})=f_{00}+h\,\nabla f^\top\mathbf{a}
+\tfrac12 h^2\,\mathbf{a}^\top H_f\,\mathbf{a}+\mathcal{O}(h^3),
\]
with \(\mathbf{a}\in\{\mathbf{u},\mathbf{v},\mathbf{u}+\mathbf{v}\}\), we obtain
\begin{align*}
f_{\mathbf{uv}}^{+} - f_{\mathbf u}^{+} - f_{\mathbf v}^{+} + f_{00}
&= \tfrac12 h^2\!\left[(\mathbf{u}+\mathbf{v})^\top H_f(\mathbf{u}+\mathbf{v})
   - \mathbf{u}^\top H_f \mathbf{u} - \mathbf{v}^\top H_f \mathbf{v}\right]\notag\\
&\quad + \mathcal O(h^3) \\
&= h^2\,\mathbf{u}^\top H_f\,\mathbf{v} + \mathcal O(h^3), \notag
\end{align*}
hence \(D_{\mathbf{uv}}^{(+)}=\mathbf{u}^\top H_f\,\mathbf{v}+\mathcal{O}(h)\).
Replacing \(h\!\to\!-h\) gives \(D_{\mathbf{uv}}^{(-)}=\mathbf{u}^\top H_f\,\mathbf{v}+\mathcal{O}(h)\).
Averaging cancels odd powers:
\begin{equation}
  D_{\mathbf{uv}}^{(\mathrm{c})}
  = \mathbf{u}^{\!\top}H_f(x_\Omega)\,\mathbf{v} \;+\; \mathcal{O}(h^{2}).
\end{equation}

The off-diagonal entry of the \(2\times 2\) shape operator in the \((\mathbf{u}, \mathbf{v})\) frame is
\[
  S_{12}(x_\Omega)
  = \frac{\mathbf{u}^{\!\top}H_f(x_\Omega)\,\mathbf{v}}{\|\nabla f(x_\Omega)\|}.
\]
Using the finite-difference approximation gives the off-diagonal Weingarten loss
\begin{equation}
  \bigl|S_{12}(x_\Omega)\bigr|
  \;\approx\;
  \frac{\bigl|D_{\mathbf{uv}}^{(\mathrm{c})}(x_\Omega)\bigr|}{\|\nabla f(x_\Omega)\|}
  \;+\;\mathcal{O}(h^{2}),
\end{equation}
and the quantity used in the squared loss becomes
\begin{equation}
  S_{12}^{2}(x_\Omega)
  \;\approx\;
  \frac{\bigl(D_{\mathbf{uv}}^{(\mathrm{c})}(x_\Omega)\bigr)^{2}}
       {\|\nabla f(x_\Omega)\|^{2}}
  \;+\;\mathcal{O}(h^{2}).
\end{equation}

In practice we may omit the division by $\|\nabla f(x_\Omega)\|$ and minimize $\lvert D_{\mathbf{uv}}^{(\mathrm{c})}\rvert$ directly: because the Eikonal loss enforces $\|\nabla f\|\!\approx\!1$ in the shell, both definitions coincide up to a small scale factor. 
This symmetric stencil converts the curvature-gap loss into a first-order computation.

\subsection{Off-Diagonal Weingarten Loss via Auto-Diff}
\label{subsec:second_order_odw}

For applications that favor numerical precision we exploit reverse-mode auto-diff to obtain a Hessian–vector product
\[
\mathbf u^{\!\top}H_f\,\mathbf v
      =\mathbf u^{\!\top}\nabla_{\!x}\!\bigl[(\nabla_{\!x}f)\!\cdot\!\mathbf v\bigr].
\]
From a shell point \(x_\Omega\) with orthonormal tangents
\(\mathbf u,\mathbf v\) (Sec.~\ref{subsec:first_order_odw})
we first run reverse mode on \(f(x_\Omega)\) to obtain the gradient
\(\mathbf g=\nabla f(x_\Omega)\).
Treating the scalar
\(g_{\mathbf v}=\mathbf g^{\!\top}\mathbf v\)
as a new objective and invoking reverse mode a second time differentiates
\(g_{\mathbf v}\) with respect to \(x_\Omega\); the result is the
Hessian–vector product \(H_f(x_\Omega)\mathbf v\).
An inner product with \(\mathbf u\) then yields
\(\mathbf u^{\!\top}H_f(x_\Omega)\mathbf v\),
and dividing by \(\|\nabla f(x_\Omega)\|\) gives the exact off-diagonal entry \(S_{12}(x_\Omega)\).

The first backward sweep delivers \(\nabla f\); a second sweep differentiates the scalar inner product \((\nabla f)\!\cdot\!\mathbf v\) with respect to the input coordinates, giving the exact mixed second derivative without materializing the full \(3{\times}3\) Hessian. 
This doubles the backward sweeps for the Weingarten loss samples and avoids the prohibitive memory footprint of full Hessian assembly. 

This Hessian–vector product therefore requires only two reverse-mode sweeps and stores no matrix entries, so its cost grows linearly with the number of shell samples and is independent of the ambient dimension. Dividing by \(\|\nabla f(x_\Omega)\|\) provides the exact \(S_{12}(x_\Omega)\) value used in the off-diagonal loss. Although more expensive than the first-order finite-difference loss, this second-order route offers higher accuracy and serves as our reference implementation for ablation studies.

\begin{figure}[t]
    \centering
    \includegraphics[width=0.98\linewidth]{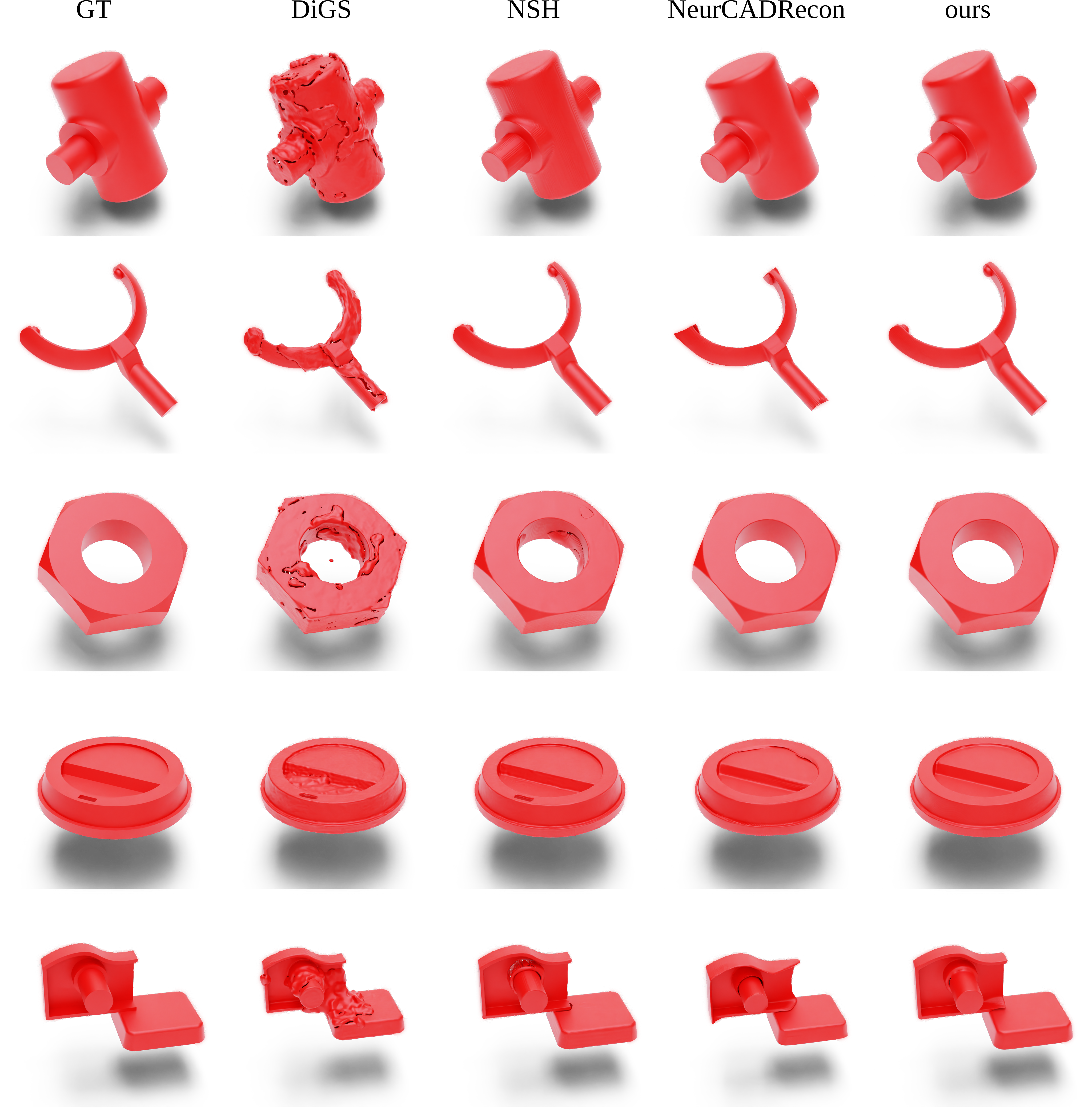}
    \caption{Comparison with state-of-the-art surface reconstruction. Our method and NeurCADRecon demonstrate superior performance in recovering clean, complete, and geometrically accurate surfaces, benefiting from curvature-aware regularization that encourages developability. Notably, our approach achieves comparable reconstruction quality while requiring only half the training time of NeurCADRecon.}

    \label{fig:collage_2}
\end{figure}

\section{Experiments} \label{sec:experiments}

In this section, we first describe the implementation details and evaluation metrics used in our validation. We then present a comprehensive comparison of our proposed off-diagonal Weingarten based loss functions (ODW-AD and ODW-FD) against three baseline methods-DiGS~\cite{BenShabat2021DiGSD}, NSH~\cite{Wang2023NeuralSingularHessianIN}, and the state-of-the-art NeurCADRecon ~\cite{Dong2024NeurCADReconNR}—on two subsets of the ABC dataset~\cite{Koch_2019_CVPR}. Finally, we analyze convergence behavior and runtime performance.

\subsection{Experimental Setup}

\paragraph{Datasets} All experiments are conducted on two 100‐example subsets of the ABC dataset~\cite{Koch_2019_CVPR}. The first subset (1\,MB set) is a pseudo‐random selection of 100 CAD models (each $\approx$ 1\,MB in file size), while the second subset (5\,MB set) comprises 100 more complex human‐chosen models (between 3.5\,MB and 9.5\,MB, in average $\approx$ 5\,MB) selected for their clean topology and well‐defined features. For each mesh, we uniformly sample a total of 30,000 points to form the input point cloud.
This preprocessing strategy ensures that all methods receive identical input distributions. After that, at each iteration 20,000 points are randomly chosen from the 30,000 points.
Additional 20,000 non-manifold points are drawn via spatial sampling within the mesh’s bounding volume.

\paragraph{Methods and Experimental Setup} We compare two variants of our off-diagonal loss: ODW-AD (Auto-Diff) and ODW-FD (Finite-Diff) against three baselines: DiGS\cite{BenShabat2021DiGSD}, Neural-Sin\-gu\-lar-Hessian (NSH)~\cite{Wang2023NeuralSingularHessianIN} and NeurCADRecon (NCR)~\cite{Dong2024NeurCADReconNR}. All methods have been implemented within a unified comparison framework to eliminate discrepancies arising from differing network architectures or training pipelines.
Each method employs a \SIREN MLP architecture~\cite{Sitzmann2020ImplicitNR} with four hidden layers of 256 units and sine activations. All models use a standard \SIREN initialization, except DiGS, which uses the MFGI initialization. Training is performed using the Adam optimizer~\cite{kingma2017adam} with a learning rate of $5\times10^{-5}$ for up to 10,000 iterations. Early stopping is triggered if the Chamfer-Distance metric plateaus for 1500 consecutive iterations. All other hyperparameters, such as weight coefficients for curvature or normal losses, and other method specific parameters follow the settings from the original implementations.

\paragraph{Hardware} Runtime and convergence experiments on the 1MB set are performed on an NVIDIA H100 GPU. Additional qualitative and quantitative experiments (e.g., on the 5\,MB set) are run on NVIDIA A100 and T4 GPUs, but only H100 results are included in the time‐comparison analysis. All machines are equipped with at least 32GB of RAM.

\subsection{Evaluation Metrics}

\paragraph{Quantitative Metrics }
We evaluate reconstruction accuracy using three standard metrics. Chamfer Distance (CD), scaled by $10^{3}$, which measures the similarity between two surfaces; lower values indicate better accuracy. F1 Score (F1) is the harmonic mean of precision and recall, computed using a distance threshold of $5\times10^{-3}$ between the predicted and ground-truth point sets. The results are scaled by $10^{2}$, with higher values indicating a better overlap. Normal Consistency (NC) captures the mean cosine similarity between predicted and ground-truth normals, scaled by $10^{2}$; higher values denote better alignment. For each of these metrics, we report the mean and standard deviation across all 100 shapes in each data subset. Additionally, for the 1MB subset, we include additional efficiency measures: GPU VRAM usage (in GB), Time per iteration (in milliseconds), and Best Iteration achieved during evaluation---the iteration count at which the best CD metric is reached. We utilize the time measurements and best iteration count to calculate the convergence time.

\subsection{Qualitative Results}
Beyond aggregate scores, the spatial distribution of reconstruction error is inspected in Figure~\ref{fig:hausdorff} on a Hausdorff-distance heat map. Our method markedly reduces peak deviations and suppresses local error spikes compared with the baselines.  Other qualitative results are shown in Figure~\ref{fig:teaser} and in the collection in Figure~\ref{fig:collage_1}. We also compare our method with other state-of-the-art reconstruction methods, depicted in Figure~\ref{fig:collage_2}, which shows our approach achieves comparable reconstruction quality with NeurCADRecon~\cite{Dong2024NeurCADReconNR}.

\begin{figure}[t]
    \centering    \includegraphics[width=0.99\linewidth]{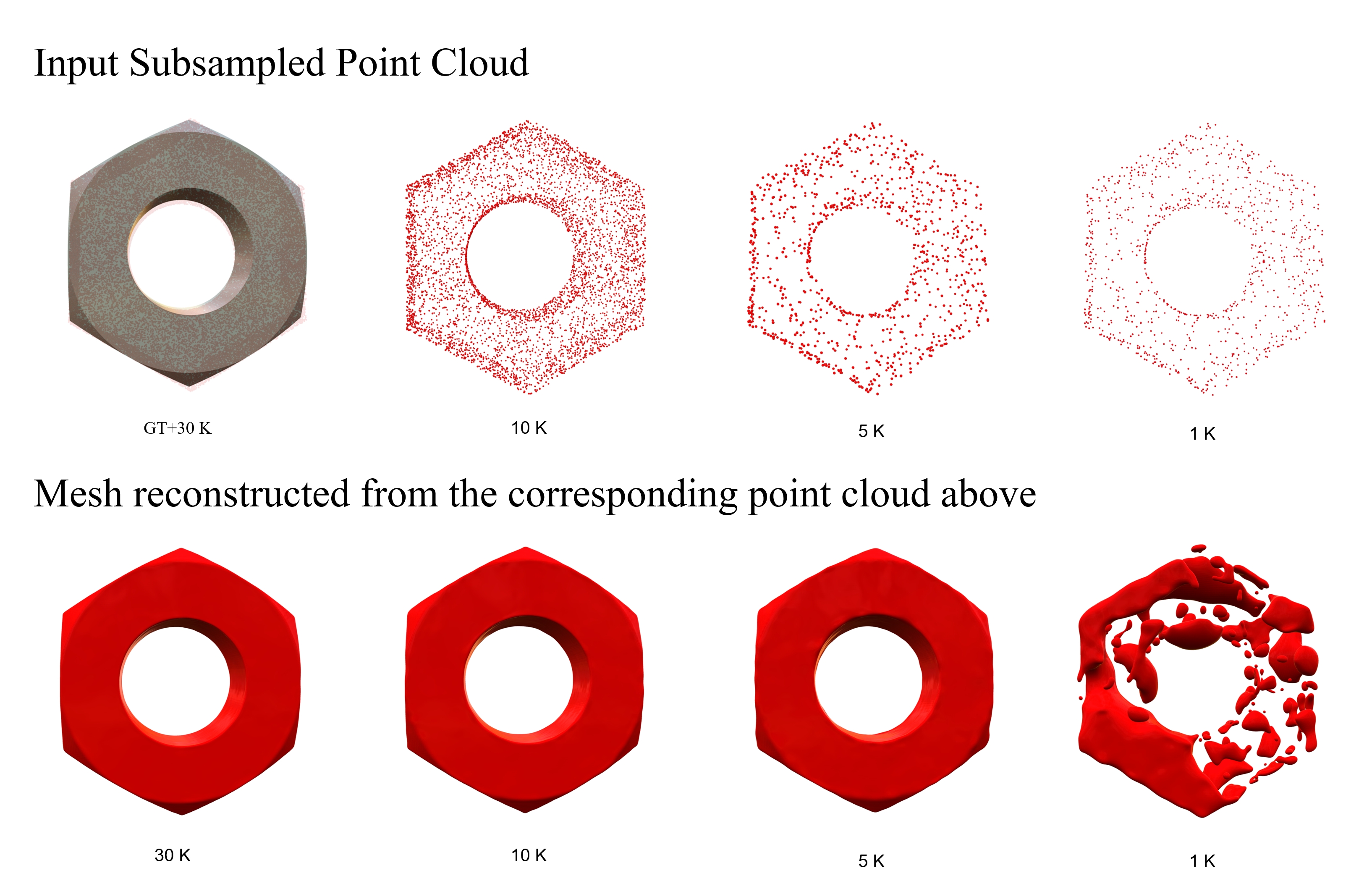}
    \caption{Qualitative results on point clouds with varying levels of sparsity. We use the original input with 30k points sampled from the ground truth mesh, and generate subsampled versions with 10k, 5k, and 1k points. The top row shows the input point clouds, and the bottom row shows the corresponding reconstructed meshes.}
    \label{fig:sparcity}
\end{figure}

\begin{figure}[t]
    \centering
    \includegraphics[width=0.99\linewidth]{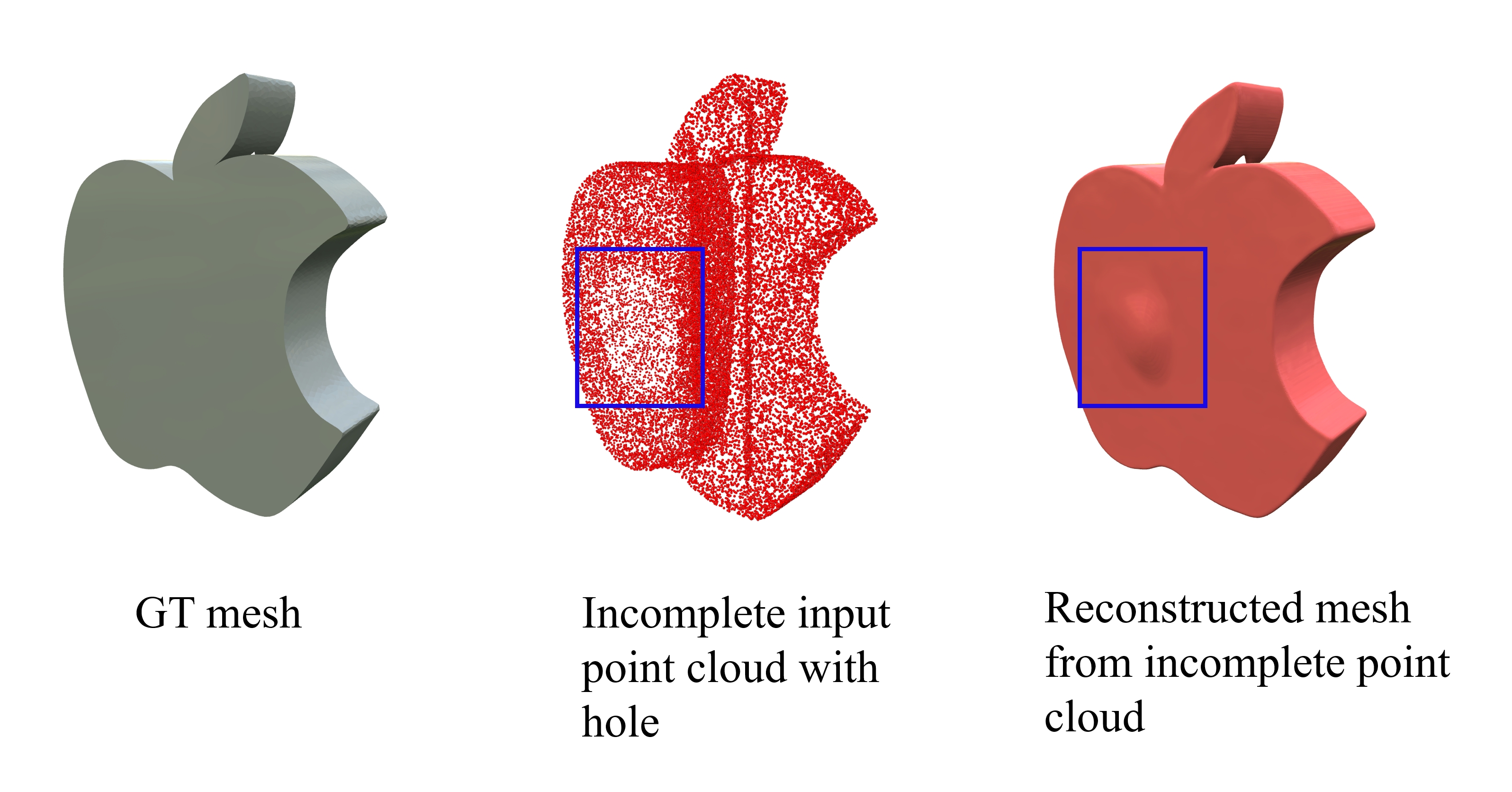}
    \caption{Using incomplete point cloud as input, our method is able to plausibly fill in missing planar regions while preserving global surface continuity and topological correctness.}
    \label{fig:incomplete}
\end{figure}

\begin{figure*}[t]
    \centering
    \includegraphics[width=0.99\linewidth]{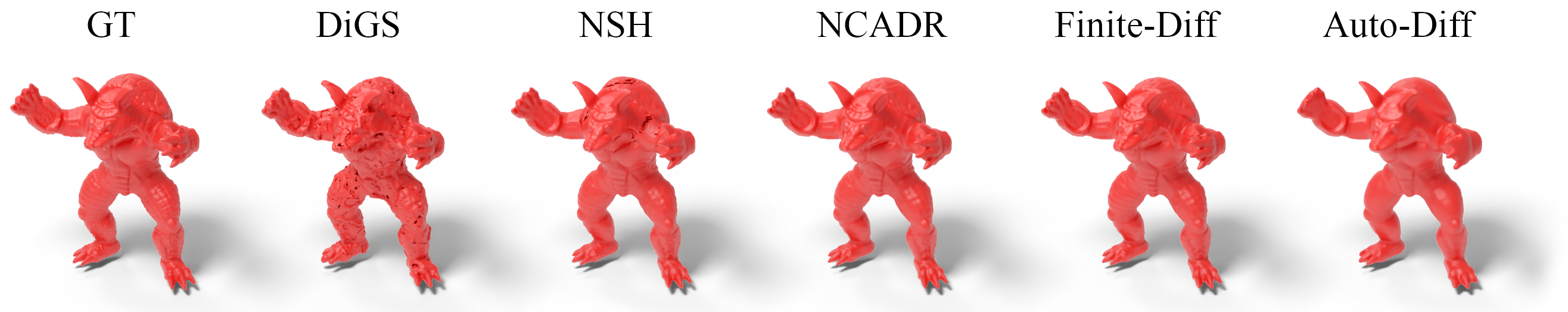}
    \caption{Reconstruction results on a non-CAD shape (Armadillo). Our method and NeurCADRecon achieve smooth and complete surfaces with fewer artifacts than DiGS and NSH. While fine details are slightly smoothed, our approach offers a good balance between quality and efficiency, with only half the training time of NeurCADRecon.}
    \label{fig:armadillo}
\end{figure*}

\subsection{Input Data Experiments}

\paragraph{Data Sparsity} To assess the robustness of our method under reduced sampling density with sparse input, we simulate sparse inputs by randomly subsampling 10k, 5k, and 1k of the originally sampled 30k point cloud data. This scenario mimics real-world settings such as thin-walled plates, tubes, or surfaces with large inter-point gaps, where predicting accurate SDF values becomes challenging.

As shown in both Figure~\ref{fig:sparcity} and quantitative Table~\ref{tab:sparse} evaluations, our method successfully reconstructs geometry of comparable quality to that obtained from the full-resolution input. Despite the significant reduction in point density, the reconstructed shapes remain smooth, topologically correct, and well-aligned with the ground truth, without introducing spurious genus or holes.

 Our method maintains high-fidelity reconstruction at 10k and 5k, while severe degradation appears at 1k due to extreme sparsity.
 
\begin{table}[b]
  \centering
  \caption{Quantitative results on sparse point-cloud data with 10k, 5k, and 1k points sub-sampled from the original 30k-point cloud generated from the ground-truth mesh.}
  \label{tab:sparse}
  \small
  \setlength{\tabcolsep}{3pt}
  \renewcommand{\arraystretch}{1.1}
  \begin{tabular*}{\linewidth}{@{\extracolsep{\fill}}lccc}
    \toprule
    \textbf{Input Size} & {NC~$\uparrow$}  & \textbf{$\mathrm{CD}_{L_1}$} & {F1~$\uparrow$}  \\
    \midrule
    30k & 0.9936044 & 0.0024629 & 0.9693 \\
    10k & 0.9926978 & 0.0024399 & 0.970385 \\
    5k  & 0.9917300 & 0.0024912 & 0.9693145 \\
    1k  & 0.7255813 & 0.0143693 & 0.3399991 \\
    \bottomrule
  \end{tabular*}
\end{table}

\paragraph{Incomplete Point Clouds}
To evaluate robustness to missing regions, a circular patch on the surface is excised from each input cloud while keeping the remaining samples unchanged.  
As reported in Table~\ref{tab:incomplete} and Figure~\ref{fig:incomplete}, the Weingarten loss-based reconstruction degrades gracefully: Chamfer distance increases by only \(64\,\%\) and normal consistency drops by \(0.7\,\%\), yet global topology and salient features are preserved, however, a visible bump remains.

\begin{table}[b]
  \centering
  \setlength{\tabcolsep}{3pt}
  \renewcommand{\arraystretch}{1.1}
  \small
  \caption{Quantitative comparison on incomplete data with hole for apple shape.}
  \label{tab:incomplete}
  \begin{tabular*}{\linewidth}{@{\extracolsep{\fill}}lccc}
    \toprule
    {input} & {NC~$\uparrow$}  & {CD$_{L_1}$~$\downarrow$}  & {F1~$\uparrow$}  \\
    \midrule
    Full & 0.9897844 & 0.0029810 & 0.9334532 \\
    Hole & 0.9824422 & 0.0048818 & 0.5045011 \\
    \bottomrule
  \end{tabular*}
\end{table}

\paragraph{Generalization to Non-CAD Shapes} Our method is specifically designed for CAD data, where the Weingarten loss discourages the reconstructed surface to be locally warped and twisted. To understand how this inductive bias affects performance on general, non-CAD geometry, we evaluate our model on the standard Armadillo shape, which contains rich organic, but also high frequency, details.

As shown in Figure~\ref{fig:armadillo} and Table~\ref{tab:armadillo}, our method produces smooth and topologically consistent reconstructions, comparable to those generated by NeurCADRecon~\cite{Dong2024NeurCADReconNR}, but at approximately half the training time. Compared to DiGS~\cite{BenShabat2021DiGSD} and Neural-Singular-Hessian (NSH)~\cite{Wang2023NeuralSingularHessianIN}, both our method and NeurCADRecon yield superior global shape recovery, exhibiting fewer holes and more coherent topology. However, we observe that fine surface details are partially smoothed out—an expected consequence of the curvature regularization, which favors larger planar or uniformly curved regions.

This experiment illustrates that, while our method is tailored for CAD reconstruction, it remains robust on general shapes and offers a favorable trade-off between geometric fidelity, smoothness, and computational efficiency.

\begin{table}[b]
  \centering
  \setlength{\tabcolsep}{3pt}
  \renewcommand{\arraystretch}{1.1}
  \small
  \caption{Quantitative results on armadillo reconstruction with different methods.}
  \label{tab:armadillo}
  \begin{tabular*}{\linewidth}{@{\extracolsep{\fill}}lcccc}
    \toprule
    {input}           & {NC~$\uparrow$}      & {CD$_{L_1}$~$\downarrow$}  & {F1~$\uparrow$}      & {time (s)} \\
    \midrule
    Our (ODW-AD)         & 0.9726746       & 0.0025855           & 0.9724751       & 877.15            \\
    Our (ODW-FD)             & 0.9799343       & 0.0023597           & 0.9877244       & 915.50            \\
    NeurCADRecon             & 0.9789649       & 0.0022634           & 0.9898299       & 1891.00           \\
    NSH                      & 0.9615119       & 0.0056421           & 0.8860692       & 231.30            \\
    DiGS                     & 0.9337240       & 0.0045840           & 0.6731318       & 714.34            \\
    \bottomrule
  \end{tabular*}
\end{table}

\subsection{Ablation Studies}
\paragraph{Ablation of Weingarten Loss Weight} To assess the impact of our proposed off-diagonal Weingarten loss regularization, we conduct an ablation study by varying the weight $\lambda_{\text{ODW}}$ applied to the loss term. Specifically, we test settings with $\lambda_{\text{ODW}} \in \{0.1, 1, 10, 100\}$ while keeping all other parameters fixed.

\begin{figure*}[t]
    \centering
    \includegraphics[width=0.99\linewidth]{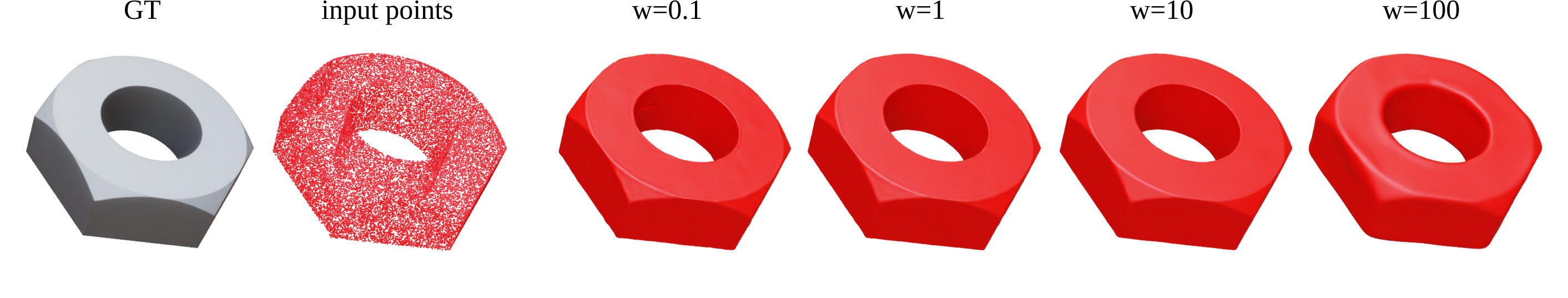}
    \caption{We compare the reconstructed shape with different Weingarten-loss weight. From our result, our method consistently produces smooth and structurally faithful reconstructions across a wide range of weights.}
    \label{fig:ablation_weights}
\end{figure*}

As shown in the quantitative results in Table~\ref{tab:ablation_weights} and qualitative comparisons in Figure~\ref{fig:ablation_weights}, our method consistently produces smooth and structurally faithful reconstructions across a wide range of weights. 
Importantly, when the curvature term is removed entirely ($\lambda_{\text{curv}} = 0$), the reconstruction quality drops significantly-exhibiting surface noise, distorted geometry, or incorrect topology in challenging regions. 
These observations confirm that even a small weight on the curvature loss is sufficient to improve reconstruction quality. The results highlight the importance of our Weingarten loss formulation in guiding the implicit surface toward CAD-compliant characteristics.

\begin{table}[b]
  \centering
  \setlength{\tabcolsep}{3pt}
  \renewcommand{\arraystretch}{1.1}
  \small
  \caption{Quantitative results with different weight settings.}
  \label{tab:ablation_weights}
  \begin{tabular*}{\linewidth}{@{\extracolsep{\fill}}lccc}
    \toprule
    {}    & {NC~$\uparrow$}      & {CD$_{L_1}$~$\downarrow$}  & {F1~$\uparrow$}  \\
    \midrule
    Weight = 100      & 0.9776159       & 0.0035313           & 0.854874  \\
    Weight = 10       & 0.9939355       & 0.0024900           & 0.969770  \\
    Weight = 1        & 0.9940296       & 0.0024305           & 0.9708199 \\
    Weight = 0.1      & 0.9920184       & 0.0025440           & 0.9596963 \\
    \bottomrule
  \end{tabular*}
\end{table}

\paragraph{Comparison of Finite-Diff and Auto-Diff} Our method supports two implementations of the curvature Weingarten loss: an \textit{Auto-Diff} version that computes the mixed second derivative analytically via backward-mode auto-differentiation, and a \textit{Finite-Diff} version that approximates the same quantity using a symmetric finite-difference stencil.

\begin{figure}[t]
    \centering
    \includegraphics[width=0.99\linewidth]{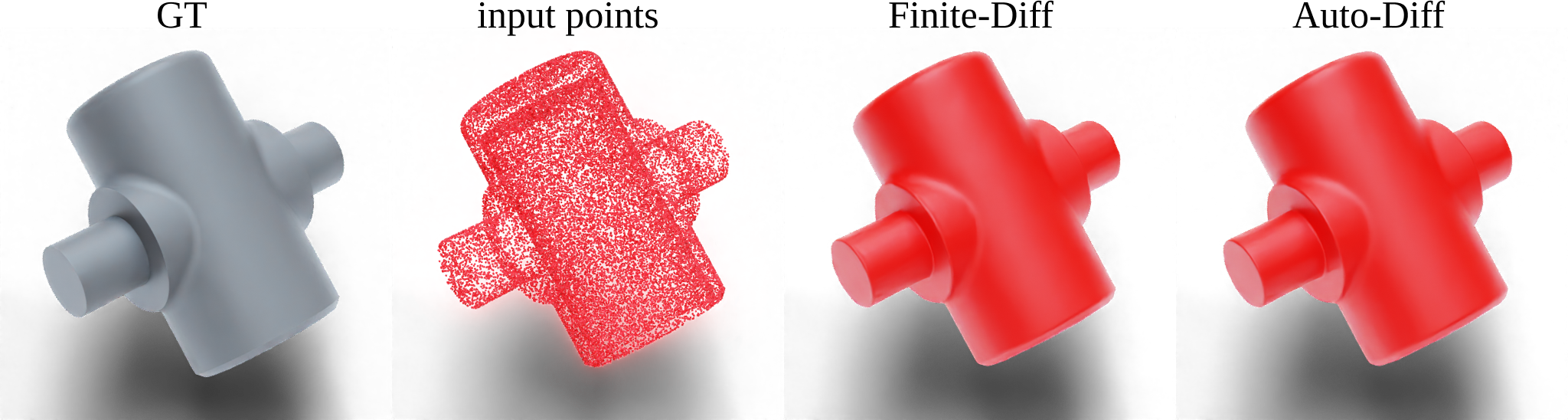}
    \caption{Ablation between Auto-Diff and Finite-Diff versions of our Weingarten loss. Both methods produce equally high-quality reconstructions, validating the effectiveness of the finite-difference approximation.}
    \label{fig:ablation_finite}
\end{figure}

To compare these alternatives, we evaluate both their reconstruction quality and runtime performance. As shown in Table~\ref{tab:ablation_odw}, Figure~\ref{fig:ablation_finite}, the two variants produce nearly identical reconstruction accuracy, suggesting that the finite-difference approximation is sufficiently precise when the neighborhood resolution we choose is small. This validates the use of our discrete formulation as a numerically stable and effective off-diagonal Weingarten loss.

In terms of runtime, we observe comparable training speeds between the two variants. While the Auto-Diff approach incurs additional backward passes 
(e.g., two extra reverse sweeps per batch), the Finite-Diff version relies on multiple forward evaluations (six additional SDF queries for the symmetric stencil; seven total including \(f_{00}\) if it is not reused). 
On networks of our scale, the difference in wall-clock performance remains negligible, further supporting the practicality of our lightweight first-order implementation.

\begin{table}[b]
  \centering
  \setlength{\tabcolsep}{3pt}
  \renewcommand{\arraystretch}{1.1}
  \small
  \caption{Comparison between Auto-Diff and Finite-Diff methods.}
  \label{tab:ablation_odw}
  \begin{tabular*}{\linewidth}{@{\extracolsep{\fill}}lcccc}
    \toprule
    {}     & {NC~$\uparrow$}     & {CD$_{L_1}$~$\downarrow$}  & {F1~$\uparrow$}        & {time (s)} \\
    \midrule
    ODW-AD          & 0.9944662       & 0.0029025           & 0.9127199         & 875.19 \\
    ODW-FD        & 0.9939950       & 0.0029577           & 0.9105444         & 909.76 \\
    \bottomrule
  \end{tabular*}
\end{table}

\begin{table}[t]
  \centering
  \small
  \caption{Comparison of iteration time (ms), convergence time (s)—computed as mean iteration time $\times$ number of iterations, scaled to seconds—and GPU memory usage. Within each column, the best is \underline{\textbf{bold and underlined}}, second-best is \textbf{bold}.}
  \label{tab:time_quant_eval}
  \renewcommand{\arraystretch}{0.9}
  \begin{tabular*}{\linewidth}{@{\extracolsep{\fill}}lccccc}
      \toprule
      & \multicolumn{2}{c}{Iter.\ time (ms)} & iter & time (s) & (GB) \\
      \cmidrule(lr){2-3}
      & mean & std. & mean & mean & mean \\
      \midrule
      DiGS~\cite{BenShabat2021DiGSD}              & 4.14 & \textbf{0.03} & 8769 & 363.04 & 3.89 \\
      NSH~\cite{Wang2023NeuralSingularHessianIN}  & 6.11 & 0.06 & 8249 & 504.01 & 6.10 \\
      NCR~\cite{Dong2024NeurCADReconNR}           & 4.11 & \textbf{\underline{0.02}} & 8129 & 334.10 & 6.16 \\
      \midrule
      \textbf{ODW‐AD}                           & \underline{\textbf{2.50}} & \textbf{0.03} & \textbf{7552} & \textbf{188.80} & \textbf{\underline{3.46}} \\
      \textbf{ODW-FD}                           & \textbf{3.09} & 0.05 & \underline{\textbf{5519}} & \textbf{\underline{170.54}} & \textbf{3.70} \\
      \bottomrule
  \end{tabular*}
\end{table}

\begin{table*}[!t]
\centering
\small
\caption{Quantitative results on the ABC dataset~\cite{Koch_2019_CVPR}. Evaluation is conducted on two resolution subsets (1\,MB and 5\,MB) using three metrics: Normal Consistency (NC), Chamfer Distance (CD), and F1 score (F1). For each metric's mean value, the best result is \underline{\textbf{bold and underlined}}, and the second-best is \textbf{bold}. NC and F1 $\times 10^2$ and CD $\times 10^3$ }
\label{tab:abc}
\renewcommand{\arraystretch}{0.9}
\begin{tabular*}{\textwidth}{@{\extracolsep{\fill}}l|cccccc|cccccc} 
\toprule
\multicolumn{1}{c|}{} & \multicolumn{6}{c|}{1\,MB set} & \multicolumn{6}{c}{5\,MB set} \\
\cmidrule{2-13}
\multicolumn{1}{c|}{} & \multicolumn{2}{c}{NC~$\uparrow$} & \multicolumn{2}{c}{CD~$\downarrow$} & \multicolumn{2}{c|}{F1~$\uparrow$} & \multicolumn{2}{c}{NC~$\uparrow$} & \multicolumn{2}{c}{CD~$\downarrow$} & \multicolumn{2}{c}{F1~$\uparrow$} \\
\multicolumn{1}{c|}{} & mean & std. & mean & std. & mean & std. & mean & std. & mean & std. & mean & std. \\ 
\midrule
DiGS~\footnotesize{\cite{BenShabat2021DiGSD}}
    & 92.30 & 5.84 & 7.21  & 4.20 & 47.91 & 30.25
    & 94.01 & 5.41 & 8.47 & 6.91 & 62.63 & 29.47 \\

NSH~\footnotesize{\cite{Wang2023NeuralSingularHessianIN}}
    & 94.84 & 6.09 & 6.73  & 6.00 & 64.94 & 34.32
    & 97.44 & 2.54 & 5.27 & 4.85 & 86.67 & 14.46 \\

NCR~\footnotesize{\cite{Dong2024NeurCADReconNR}}
    & \underline{\textbf{95.43}} & \underline{\textbf{4.92}} & \textbf{3.92} & \textbf{3.72} & \underline{\textbf{87.74}} & \textbf{16.33}
    & \underline{\textbf{97.59}} & \underline{\textbf{2.35}} & \textbf{4.99} & \underline{\textbf{4.17}} & \underline{\textbf{88.29}} & 15.65 \\

\midrule
\textbf{ODW‐AD}
    & 93.79 & 6.80 & 4.59 & 3.98 & 83.80 & 18.28
    & 97.23 & 2.96 & 5.27 & 4.60 & \textbf{86.82} & \textbf{\underline{13.81}} \\

\textbf{ODW-FD}
    & \textbf{94.69} & \textbf{5.41} & \underline{\textbf{3.84}} & \textbf{\underline{3.44}} & \textbf{87.65} & \textbf{\underline{16.02}}
    & \textbf{97.46} & \textbf{2.48} & \underline{\textbf{4.93}} & \textbf{4.38} & 86.61 & \textbf{13.92} \\
\bottomrule
\end{tabular*}
\end{table*}

\subsection{Quantitative Results}
\label{sec:validation}

\begin{figure*}[t]
    \centering
    \includegraphics[width=0.999\linewidth]{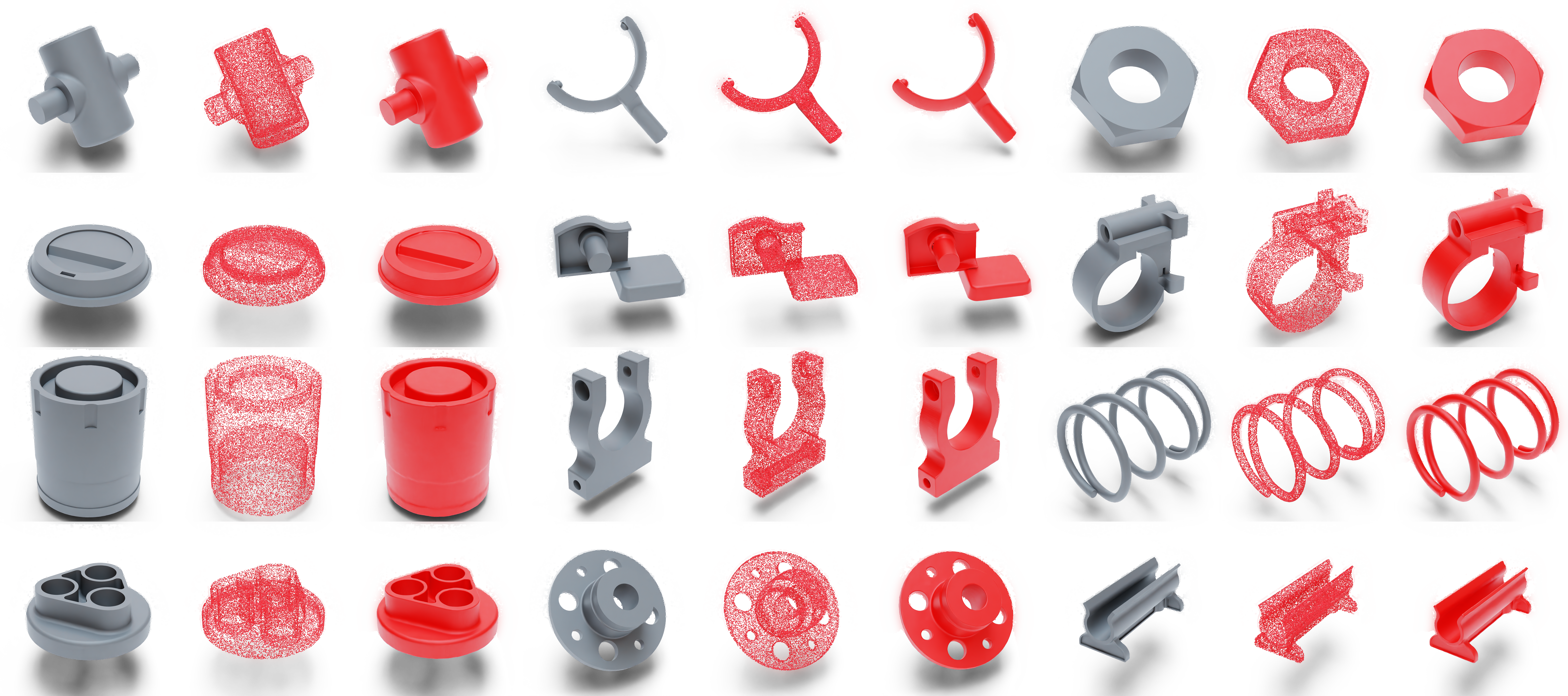}
    \caption{Results of our method, each set shows the ground truth, the used point clouds, and the rendering of our reconstruction. }
    \label{fig:collage_1}
\end{figure*}

\paragraph{Human-selected Examples}
On the cleaner 5\,MB subset, see Table \ref{tab:abc}, both Weingarten variants exhibit strong performance across all metrics. ODW-FD achieves the lowest Chamfer Distance (CD), outperforming NeurCADRecon, while still keeping the Normal Consistency (NC) value very competitive. ODW‐AD achieves an F1 score only second to NeurCADRecon, while keeping the variance lower. These results highlight the ability of both Weingarten loss methods to generalize well on well-curated data, with ODW-FD showing the most consistent geometric accuracy, and ODW‐AD providing the highest detection fidelity. Figure~\ref{fig:collage_1} depicts a few selected models of the 5\,MB set. 

\paragraph{Pseudo-Random Examples}
In the more varied 1\,MB subset, see Table \ref{tab:abc}, where input meshes are less structured, ODW-FD leads in geometric accuracy with the lowest CD and a strong NC, closely trailing NeurCADRecon. F1 scores remain competitive, with ODW-FD and ODW‐AD closely rivaling NeurCADRecon. These findings demonstrate that even with varied input point clouds, both ODW-FD and ODW‐AD deliver reconstructions that are either on par with or superior to NCR, and significantly outperform DiGS and NSH across all metrics.

\paragraph{Efficiency}
In terms of computational efficiency, see Table \ref{tab:time_quant_eval}, both of our methods clearly outperform the baselines in terms of iteration time ($\approx2\times$ speedup over NCR). The difference is even more apparent when considering the overall training convergence time ($\approx2\times$ speedup over NCR). ODW‐AD offers a strong trade-off with the second-lowest iteration count and runtime, while consuming modest GPU memory. Compared to NeurCADRecon's long runtime and high memory footprint, both Weingarten methods are significantly more efficient. This demonstrates that ODW-FD and ODW‐AD not only offer high-quality reconstructions but do so with favorable resource demands, making them well suited for scalable and time-sensitive applications.

\section{Discussion and Conclusions}

\paragraph{Discussion} 
The proposed off-diagonal Weingarten loss term minimizes the expectation $\mathcal L_{\text{ODW}}=\mathbb E_\theta[S_{12}^{2}]\propto(\kappa_{2}-\kappa_{1})^{2}$, so gradient descent drives the two principal curvatures to coincide.  Where one curvature is already zero (parabolic zones) or of opposite sign (hyperbolic zones) the only viable limit is $\kappa_{1}=\kappa_{2}=0$, hence those patches relax to planes.  Where both curvatures share a sign (elliptic zones) equality produces an umbilic state and the patch rounds into a spherical cap whose radius is fixed by the data term.  In this way the loss selectively flattens developable and saddle parts while regularizing doubly-curved areas without erasing their overall shape.

Because the Weingarten loss penalty acts only on the \textit{curvature difference}, it complements rather than competes with the Dirichlet and Eikonal constraints: the Dirichlet term fits the sampled points exactly, the Eikonal term keeps the normal lines straight, and the Weingarten loss term eliminates the residual tangential “twist” that is unconstrained by first-order information.  

Compared to the Gaussian-curvature loss introduced in previous work \cite{Dong2024NeurCADReconNR}, we showed in our results that, under identical shapes and training settings, our Weingarten loss delivers essentially the same reconstruction accuracy and boundary sharpness while requiring only one Hessian–vector product per sample instead of the multiple second-order evaluations and quadratic combination needed for the Gaussian term, resulting in a markedly lower compute and memory footprint during training.

\paragraph{Limitations and Future Work} 

The off-diagonal Weingarten loss is lightweight and effective, yet several boundaries remain.  Because it equalizes the two principal curvatures, it assigns a non-zero penalty to developable but curved primitives such as cylinders and cones; with a balanced weight the data term keeps the correct radius or slope, but an overly strong weight can nudge these patches toward planar or spherical limits.  The regularizer acts purely locally in the off-surface shell and carries no notion of long-range patch coherence, so global fairness constraints—keeping opposite faces of a thin sheet parallel, for example—are not guaranteed.

The exact Hessian-vector implementation also doubles the backward pass for Weingarten samples, so the finite-difference stencil remains preferable on memory-constrained hardware, with a truncation error \(\mathcal{O}(h^2)\) for the symmetric stencil (or \(\mathcal{O}(h)\) if only the one-sided stencil is used).
Finally, the Weingarten loss has no explicit mechanism for preserving sharp feature lines: narrow chamfers are retained only when they are faithfully sampled and reproduced by the decoder, suggesting that an edge-aware weighting scheme is a useful direction for future work.

Our off-diagonal Weingarten loss treats curvature reduction purely locally; an immediate extension is to learn an adaptive weight $\lambda_{\text{ODW}}$ that depends on local feature probability, so planar zones flatten faster while highly curved zones remain expressive.
A second avenue is joint optimization with mesh extraction: coupling the Weingarten loss with differentiable DMC/dual-contouring could avoid the Marching-Cubes voxel bias.  

\paragraph{Conclusions}
We introduced the off-diagonal Weingarten loss that measures the curvature gap and needs only one Hessian-vector product, giving $\approx 2 \times$ faster convergence than full Gaussian-curvature regularization and using roughly half the GPU memory, while matching or surpassing state-of-the-art accuracy on the ABC benchmark. Two proposed implementations  (auto-diff and finite-diff) perform similarly; the finite-difference stencil is preferable when back-pro\-pa\-ga\-tion memory is scarce. The technique is drop-in for any SIREN-based reconstruction pipeline and our code is available at \url{https://flatcad.github.io/}. 

\section*{Acknowledgments}
We thank the anonymous reviewers for valuable inputs and our institutions for supporting the project. 

{
    \bibliographystyle{ACM-Reference-Format}
    \bibliography{main}  


\begin{thebibliography}{40}

%%% the Web address.

\ifx \showCODEN    \undefined \def \showCODEN     #1{\unskip}     \fi
\ifx \showDOI      \undefined \def \showDOI       #1{#1}\fi
\ifx \showISBNx    \undefined \def \showISBNx     #1{\unskip}     \fi
\ifx \showISBNxiii \undefined \def \showISBNxiii  #1{\unskip}     \fi
\ifx \showISSN     \undefined \def \showISSN      #1{\unskip}     \fi
\ifx \showLCCN     \undefined \def \showLCCN      #1{\unskip}     \fi
\ifx \shownote     \undefined \def \shownote      #1{#1}          \fi
\ifx \showarticletitle \undefined \def \showarticletitle #1{#1}   \fi
\ifx \showURL      \undefined \def \showURL       {\relax}        \fi
% invisible to TeX
\providecommand\bibfield[2]{#2}
\providecommand\bibinfo[2]{#2}
\providecommand\natexlab[1]{#1}
\providecommand\showeprint[2][]{arXiv:#2}

\bibitem[Atzmon and Lipman(2019)]%
        {Atzmon2019SALSA}
\bibfield{author}{\bibinfo{person}{Matan Atzmon} {and} \bibinfo{person}{Yaron Lipman}.} \bibinfo{year}{2019}\natexlab{}.
\newblock \showarticletitle{SAL: Sign Agnostic Learning of Shapes From Raw Data}.
\newblock \bibinfo{journal}{\emph{2020 IEEE/CVF Conference on Computer Vision and Pattern Recognition (CVPR)}} (\bibinfo{year}{2019}), \bibinfo{pages}{2562--2571}.
\newblock
\urldef\tempurl%
\url{https://api.semanticscholar.org/CorpusID:208267630}
\showURL{%
\tempurl}


\bibitem[Atzmon and Lipman(2020)]%
        {Atzmon2020SALSA}
\bibfield{author}{\bibinfo{person}{Matan Atzmon} {and} \bibinfo{person}{Yaron Lipman}.} \bibinfo{year}{2020}\natexlab{}.
\newblock \showarticletitle{SAL++: Sign Agnostic Learning with Derivatives}.
\newblock \bibinfo{journal}{\emph{ArXiv}}  \bibinfo{volume}{abs/2006.05400} (\bibinfo{year}{2020}).
\newblock
\urldef\tempurl%
\url{https://api.semanticscholar.org/CorpusID:219558695}
\showURL{%
\tempurl}


\bibitem[Ben-Shabat et~al\mbox{.}(2021)]%
        {BenShabat2021DiGSD}
\bibfield{author}{\bibinfo{person}{Yizhak Ben-Shabat}, \bibinfo{person}{Chamin Pasidu~Hewa Koneputugodage}, {and} \bibinfo{person}{Stephen Gould}.} \bibinfo{year}{2021}\natexlab{}.
\newblock \showarticletitle{DiGS : Divergence guided shape implicit neural representation for unoriented point clouds}.
\newblock \bibinfo{journal}{\emph{2022 IEEE/CVF Conference on Computer Vision and Pattern Recognition (CVPR)}} (\bibinfo{year}{2021}), \bibinfo{pages}{19301--19310}.
\newblock
\urldef\tempurl%
\url{https://api.semanticscholar.org/CorpusID:235490568}
\showURL{%
\tempurl}


\bibitem[Boulch and Marlet(2022)]%
        {Boulch2022POCOPC}
\bibfield{author}{\bibinfo{person}{Alexandre Boulch} {and} \bibinfo{person}{Renaud Marlet}.} \bibinfo{year}{2022}\natexlab{}.
\newblock \showarticletitle{POCO: Point Convolution for Surface Reconstruction}.
\newblock \bibinfo{journal}{\emph{2022 IEEE/CVF Conference on Computer Vision and Pattern Recognition (CVPR)}} (\bibinfo{year}{2022}), \bibinfo{pages}{6292--6304}.
\newblock
\urldef\tempurl%
\url{https://api.semanticscholar.org/CorpusID:245769792}
\showURL{%
\tempurl}


\bibitem[Carr et~al\mbox{.}(2001)]%
        {Carr2001ReconstructionAR}
\bibfield{author}{\bibinfo{person}{Jonathan~C. Carr}, \bibinfo{person}{Richard~K. Beatson}, \bibinfo{person}{Jon~B. Cherrie}, \bibinfo{person}{Tim~J. Mitchell}, \bibinfo{person}{W.~Richard Fright}, \bibinfo{person}{Bruce~C. McCallum}, {and} \bibinfo{person}{Tim~R. Evans}.} \bibinfo{year}{2001}\natexlab{}.
\newblock \showarticletitle{Reconstruction and representation of 3D objects with radial basis functions}.
\newblock \bibinfo{journal}{\emph{Proceedings of the 28th annual conference on Computer graphics and interactive techniques}} (\bibinfo{year}{2001}).
\newblock
\urldef\tempurl%
\url{https://api.semanticscholar.org/CorpusID:11863186}
\showURL{%
\tempurl}


\bibitem[Chabra et~al\mbox{.}(2020)]%
        {Chabra2020DeepLS}
\bibfield{author}{\bibinfo{person}{Rohan Chabra}, \bibinfo{person}{Jan~Eric Lenssen}, \bibinfo{person}{Eddy Ilg}, \bibinfo{person}{Tanner Schmidt}, \bibinfo{person}{Julian Straub}, \bibinfo{person}{S. Lovegrove}, {and} \bibinfo{person}{Richard~A. Newcombe}.} \bibinfo{year}{2020}\natexlab{}.
\newblock \showarticletitle{Deep Local Shapes: Learning Local SDF Priors for Detailed 3D Reconstruction}. In \bibinfo{booktitle}{\emph{European Conference on Computer Vision}}.
\newblock
\urldef\tempurl%
\url{https://api.semanticscholar.org/CorpusID:214623267}
\showURL{%
\tempurl}


\bibitem[Chibane et~al\mbox{.}(2020)]%
        {Chibane2020ImplicitFI}
\bibfield{author}{\bibinfo{person}{Julian Chibane}, \bibinfo{person}{Thiemo Alldieck}, {and} \bibinfo{person}{Gerard Pons-Moll}.} \bibinfo{year}{2020}\natexlab{}.
\newblock \showarticletitle{Implicit Functions in Feature Space for 3D Shape Reconstruction and Completion}.
\newblock \bibinfo{journal}{\emph{2020 IEEE/CVF Conference on Computer Vision and Pattern Recognition (CVPR)}} (\bibinfo{year}{2020}), \bibinfo{pages}{6968--6979}.
\newblock
\urldef\tempurl%
\url{https://api.semanticscholar.org/CorpusID:211817828}
\showURL{%
\tempurl}


\bibitem[Dong et~al\mbox{.}(2024)]%
        {Dong2024NeurCADReconNR}
\bibfield{author}{\bibinfo{person}{Qiujie Dong}, \bibinfo{person}{Rui Xu}, \bibinfo{person}{Pengfei Wang}, \bibinfo{person}{Shuangmin Chen}, \bibinfo{person}{Shiqing Xin}, \bibinfo{person}{Xiaohong Jia}, \bibinfo{person}{Wenping Wang}, {and} \bibinfo{person}{Changhe Tu}.} \bibinfo{year}{2024}\natexlab{}.
\newblock \showarticletitle{NeurCADRecon: Neural Representation for Reconstructing CAD Surfaces by Enforcing Zero Gaussian Curvature}.
\newblock \bibinfo{journal}{\emph{ACM Transactions on Graphics (TOG)}}  \bibinfo{volume}{43} (\bibinfo{year}{2024}), \bibinfo{pages}{1 -- 17}.
\newblock
\urldef\tempurl%
\url{https://api.semanticscholar.org/CorpusID:269293976}
\showURL{%
\tempurl}


\bibitem[Erler et~al\mbox{.}(2020)]%
        {Erler2020Points2SurfLI}
\bibfield{author}{\bibinfo{person}{Philipp Erler}, \bibinfo{person}{Paul Guerrero}, \bibinfo{person}{Stefan Ohrhallinger}, \bibinfo{person}{Niloy~Jyoti Mitra}, {and} \bibinfo{person}{Michael Wimmer}.} \bibinfo{year}{2020}\natexlab{}.
\newblock \showarticletitle{Points2Surf Learning Implicit Surfaces from Point Clouds}. In \bibinfo{booktitle}{\emph{European Conference on Computer Vision}}.
\newblock
\urldef\tempurl%
\url{https://api.semanticscholar.org/CorpusID:226298441}
\showURL{%
\tempurl}


\bibitem[Gropp et~al\mbox{.}(2020)]%
        {Gropp2020ImplicitGR}
\bibfield{author}{\bibinfo{person}{Amos Gropp}, \bibinfo{person}{Lior Yariv}, \bibinfo{person}{Niv Haim}, \bibinfo{person}{Matan Atzmon}, {and} \bibinfo{person}{Yaron Lipman}.} \bibinfo{year}{2020}\natexlab{}.
\newblock \showarticletitle{Implicit Geometric Regularization for Learning Shapes}. In \bibinfo{booktitle}{\emph{International Conference on Machine Learning}}.
\newblock
\urldef\tempurl%
\url{https://api.semanticscholar.org/CorpusID:211259068}
\showURL{%
\tempurl}


\bibitem[Hoppe et~al\mbox{.}(1992)]%
        {Hoppe1992SurfaceRF}
\bibfield{author}{\bibinfo{person}{Hugues Hoppe}, \bibinfo{person}{Tony DeRose}, \bibinfo{person}{Tom Duchamp}, \bibinfo{person}{John~Alan McDonald}, {and} \bibinfo{person}{Werner Stuetzle}.} \bibinfo{year}{1992}\natexlab{}.
\newblock \showarticletitle{Surface reconstruction from unorganized points}.
\newblock \bibinfo{journal}{\emph{Proceedings of the 19th annual conference on Computer graphics and interactive techniques}} (\bibinfo{year}{1992}).
\newblock
\urldef\tempurl%
\url{https://api.semanticscholar.org/CorpusID:1122241}
\showURL{%
\tempurl}


\bibitem[Hou et~al\mbox{.}(2022)]%
        {Hou2022IterativePS}
\bibfield{author}{\bibinfo{person}{Fei Hou}, \bibinfo{person}{Chiyu Wang}, \bibinfo{person}{Wencheng Wang}, \bibinfo{person}{Hong Qin}, \bibinfo{person}{Chen Qian}, {and} \bibinfo{person}{Ying He}.} \bibinfo{year}{2022}\natexlab{}.
\newblock \showarticletitle{Iterative poisson surface reconstruction (iPSR) for unoriented points}.
\newblock \bibinfo{journal}{\emph{ACM Transactions on Graphics (TOG)}}  \bibinfo{volume}{41} (\bibinfo{year}{2022}), \bibinfo{pages}{1 -- 13}.
\newblock
\urldef\tempurl%
\url{https://api.semanticscholar.org/CorpusID:250956751}
\showURL{%
\tempurl}


\bibitem[Huang et~al\mbox{.}(2022)]%
        {Huang2022ANG}
\bibfield{author}{\bibinfo{person}{Jiahui Huang}, \bibinfo{person}{Haoxiang Chen}, {and} \bibinfo{person}{Shihui Hu}.} \bibinfo{year}{2022}\natexlab{}.
\newblock \showarticletitle{A Neural Galerkin Solver for Accurate Surface Reconstruction}.
\newblock \bibinfo{journal}{\emph{ACM Transactions on Graphics (TOG)}}  \bibinfo{volume}{41} (\bibinfo{year}{2022}), \bibinfo{pages}{1 -- 16}.
\newblock
\urldef\tempurl%
\url{https://api.semanticscholar.org/CorpusID:254097072}
\showURL{%
\tempurl}


\bibitem[Huang et~al\mbox{.}(2019)]%
        {Huang2019VariationalIP}
\bibfield{author}{\bibinfo{person}{Zhiyang Huang}, \bibinfo{person}{Nathan~A. Carr}, {and} \bibinfo{person}{Tao Ju}.} \bibinfo{year}{2019}\natexlab{}.
\newblock \showarticletitle{Variational implicit point set surfaces}.
\newblock \bibinfo{journal}{\emph{ACM Transactions on Graphics (TOG)}}  \bibinfo{volume}{38} (\bibinfo{year}{2019}), \bibinfo{pages}{1 -- 13}.
\newblock
\urldef\tempurl%
\url{https://api.semanticscholar.org/CorpusID:196834922}
\showURL{%
\tempurl}


\bibitem[Jiang et~al\mbox{.}(2020)]%
        {Jiang2020LocalIG}
\bibfield{author}{\bibinfo{person}{Chiyu~Max Jiang}, \bibinfo{person}{Avneesh Sud}, \bibinfo{person}{Ameesh Makadia}, \bibinfo{person}{Jingwei Huang}, \bibinfo{person}{Matthias Nie{\ss}ner}, {and} \bibinfo{person}{Thomas~A. Funkhouser}.} \bibinfo{year}{2020}\natexlab{}.
\newblock \showarticletitle{Local Implicit Grid Representations for 3D Scenes}.
\newblock \bibinfo{journal}{\emph{2020 IEEE/CVF Conference on Computer Vision and Pattern Recognition (CVPR)}} (\bibinfo{year}{2020}), \bibinfo{pages}{6000--6009}.
\newblock
\urldef\tempurl%
\url{https://api.semanticscholar.org/CorpusID:214606025}
\showURL{%
\tempurl}


\bibitem[Kania et~al\mbox{.}(2020)]%
        {Kania2020UCSGNetU}
\bibfield{author}{\bibinfo{person}{Kacper Kania}, \bibinfo{person}{Maciej Zieba}, {and} \bibinfo{person}{Tomasz Kajdanowicz}.} \bibinfo{year}{2020}\natexlab{}.
\newblock \showarticletitle{UCSG-Net - Unsupervised Discovering of Constructive Solid Geometry Tree}.
\newblock \bibinfo{journal}{\emph{ArXiv}}  \bibinfo{volume}{abs/2006.09102} (\bibinfo{year}{2020}).
\newblock
\urldef\tempurl%
\url{https://api.semanticscholar.org/CorpusID:219708841}
\showURL{%
\tempurl}


\bibitem[Kazhdan et~al\mbox{.}(2020)]%
        {Kazhdan2020PoissonSR}
\bibfield{author}{\bibinfo{person}{Misha Kazhdan}, \bibinfo{person}{Ming Chuang}, \bibinfo{person}{Szymon Rusinkiewicz}, {and} \bibinfo{person}{Hugues Hoppe}.} \bibinfo{year}{2020}\natexlab{}.
\newblock \showarticletitle{Poisson Surface Reconstruction with Envelope Constraints}.
\newblock \bibinfo{journal}{\emph{Computer Graphics Forum}}  \bibinfo{volume}{39} (\bibinfo{year}{2020}).
\newblock
\urldef\tempurl%
\url{https://api.semanticscholar.org/CorpusID:220116886}
\showURL{%
\tempurl}


\bibitem[Kazhdan et~al\mbox{.}(2006)]%
        {Kazhdan2006PoissonSR}
\bibfield{author}{\bibinfo{person}{Michael~M. Kazhdan}, \bibinfo{person}{Matthew Bolitho}, {and} \bibinfo{person}{Hugues Hoppe}.} \bibinfo{year}{2006}\natexlab{}.
\newblock \showarticletitle{Poisson surface reconstruction}. In \bibinfo{booktitle}{\emph{Eurographics Symposium on Geometry Processing}}.
\newblock
\urldef\tempurl%
\url{https://api.semanticscholar.org/CorpusID:14224}
\showURL{%
\tempurl}


\bibitem[Kazhdan and Hoppe(2013)]%
        {Kazhdan2013ScreenedPS}
\bibfield{author}{\bibinfo{person}{Michael~M. Kazhdan} {and} \bibinfo{person}{Hugues Hoppe}.} \bibinfo{year}{2013}\natexlab{}.
\newblock \showarticletitle{Screened poisson surface reconstruction}.
\newblock \bibinfo{journal}{\emph{ACM Trans. Graph.}}  \bibinfo{volume}{32} (\bibinfo{year}{2013}), \bibinfo{pages}{29:1--29:13}.
\newblock
\urldef\tempurl%
\url{https://api.semanticscholar.org/CorpusID:1371704}
\showURL{%
\tempurl}


\bibitem[Kingma and Ba(2017)]%
        {kingma2017adam}
\bibfield{author}{\bibinfo{person}{Diederik~P. Kingma} {and} \bibinfo{person}{Jimmy Ba}.} \bibinfo{year}{2017}\natexlab{}.
\newblock \showarticletitle{Adam: A Method for Stochastic Optimization}.
\newblock  (\bibinfo{year}{2017}).
\newblock
\showeprint[arxiv]{1412.6980}~[cs.LG]


\bibitem[Koch et~al\mbox{.}(2019)]%
        {Koch_2019_CVPR}
\bibfield{author}{\bibinfo{person}{Sebastian Koch}, \bibinfo{person}{Albert Matveev}, \bibinfo{person}{Zhongshi Jiang}, \bibinfo{person}{Francis Williams}, \bibinfo{person}{Alexey Artemov}, \bibinfo{person}{Evgeny Burnaev}, \bibinfo{person}{Marc Alexa}, \bibinfo{person}{Denis Zorin}, {and} \bibinfo{person}{Daniele Panozzo}.} \bibinfo{year}{2019}\natexlab{}.
\newblock \showarticletitle{ABC: A Big CAD Model Dataset For Geometric Deep Learning}. In \bibinfo{booktitle}{\emph{The IEEE Conference on Computer Vision and Pattern Recognition (CVPR)}}.
\newblock


\bibitem[Li et~al\mbox{.}(2018)]%
        {Li2018SupervisedFO}
\bibfield{author}{\bibinfo{person}{Lingxiao Li}, \bibinfo{person}{Minhyuk Sung}, \bibinfo{person}{Anastasia Dubrovina}, \bibinfo{person}{L. Yi}, {and} \bibinfo{person}{Leonidas~J. Guibas}.} \bibinfo{year}{2018}\natexlab{}.
\newblock \showarticletitle{Supervised Fitting of Geometric Primitives to 3D Point Clouds}.
\newblock \bibinfo{journal}{\emph{2019 IEEE/CVF Conference on Computer Vision and Pattern Recognition (CVPR)}} (\bibinfo{year}{2018}), \bibinfo{pages}{2647--2655}.
\newblock
\urldef\tempurl%
\url{https://api.semanticscholar.org/CorpusID:53715802}
\showURL{%
\tempurl}


\bibitem[Li et~al\mbox{.}(2016)]%
        {Li2016SparseRS}
\bibfield{author}{\bibinfo{person}{Manyi Li}, \bibinfo{person}{Falai Chen}, \bibinfo{person}{Wenping Wang}, {and} \bibinfo{person}{Changhe Tu}.} \bibinfo{year}{2016}\natexlab{}.
\newblock \showarticletitle{Sparse RBF surface representations}.
\newblock \bibinfo{journal}{\emph{Comput. Aided Geom. Des.}}  \bibinfo{volume}{48} (\bibinfo{year}{2016}), \bibinfo{pages}{49--59}.
\newblock
\urldef\tempurl%
\url{https://api.semanticscholar.org/CorpusID:35948521}
\showURL{%
\tempurl}


\bibitem[Lin et~al\mbox{.}(2022)]%
        {Lin2022SurfaceRF}
\bibfield{author}{\bibinfo{person}{Siyou Lin}, \bibinfo{person}{Dong Xiao}, \bibinfo{person}{Zuoqiang Shi}, {and} \bibinfo{person}{Bin Wang}.} \bibinfo{year}{2022}\natexlab{}.
\newblock \showarticletitle{Surface Reconstruction from Point Clouds without Normals by Parametrizing the Gauss Formula}.
\newblock \bibinfo{journal}{\emph{ACM Transactions on Graphics}}  \bibinfo{volume}{42} (\bibinfo{year}{2022}), \bibinfo{pages}{1 -- 19}.
\newblock
\urldef\tempurl%
\url{https://api.semanticscholar.org/CorpusID:251257800}
\showURL{%
\tempurl}


\bibitem[Mescheder et~al\mbox{.}(2018)]%
        {Mescheder2018OccupancyNL}
\bibfield{author}{\bibinfo{person}{Lars~M. Mescheder}, \bibinfo{person}{Michael Oechsle}, \bibinfo{person}{Michael Niemeyer}, \bibinfo{person}{Sebastian Nowozin}, {and} \bibinfo{person}{Andreas Geiger}.} \bibinfo{year}{2018}\natexlab{}.
\newblock \showarticletitle{Occupancy Networks: Learning 3D Reconstruction in Function Space}.
\newblock \bibinfo{journal}{\emph{2019 IEEE/CVF Conference on Computer Vision and Pattern Recognition (CVPR)}} (\bibinfo{year}{2018}), \bibinfo{pages}{4455--4465}.
\newblock
\urldef\tempurl%
\url{https://api.semanticscholar.org/CorpusID:54465161}
\showURL{%
\tempurl}


\bibitem[M{\"u}ller et~al\mbox{.}(2022)]%
        {Mller2022InstantNG}
\bibfield{author}{\bibinfo{person}{Thomas M{\"u}ller}, \bibinfo{person}{Alex Evans}, \bibinfo{person}{Christoph Schied}, {and} \bibinfo{person}{Alexander Keller}.} \bibinfo{year}{2022}\natexlab{}.
\newblock \showarticletitle{Instant neural graphics primitives with a multiresolution hash encoding}.
\newblock \bibinfo{journal}{\emph{ACM Transactions on Graphics (TOG)}}  \bibinfo{volume}{41} (\bibinfo{year}{2022}), \bibinfo{pages}{1 -- 15}.
\newblock
\urldef\tempurl%
\url{https://api.semanticscholar.org/CorpusID:246016186}
\showURL{%
\tempurl}


\bibitem[Novello et~al\mbox{.}(2023)]%
        {Novello_2023_ICCV}
\bibfield{author}{\bibinfo{person}{Tiago Novello}, \bibinfo{person}{Vinicius da Silva}, \bibinfo{person}{Guilherme Schardong}, \bibinfo{person}{Luiz Schirmer}, \bibinfo{person}{Helio Lopes}, {and} \bibinfo{person}{Luiz Velho}.} \bibinfo{year}{2023}\natexlab{}.
\newblock \showarticletitle{Neural Implicit Surface Evolution}. In \bibinfo{booktitle}{\emph{Proceedings of the IEEE/CVF International Conference on Computer Vision (ICCV)}}. \bibinfo{pages}{14279--14289}.
\newblock


\bibitem[Park et~al\mbox{.}(2019)]%
        {Park2019DeepSDFLC}
\bibfield{author}{\bibinfo{person}{Jeong~Joon Park}, \bibinfo{person}{Peter~R. Florence}, \bibinfo{person}{Julian Straub}, \bibinfo{person}{Richard~A. Newcombe}, {and} \bibinfo{person}{S. Lovegrove}.} \bibinfo{year}{2019}\natexlab{}.
\newblock \showarticletitle{DeepSDF: Learning Continuous Signed Distance Functions for Shape Representation}.
\newblock \bibinfo{journal}{\emph{2019 IEEE/CVF Conference on Computer Vision and Pattern Recognition (CVPR)}} (\bibinfo{year}{2019}), \bibinfo{pages}{165--174}.
\newblock
\urldef\tempurl%
\url{https://api.semanticscholar.org/CorpusID:58007025}
\showURL{%
\tempurl}


\bibitem[Peng et~al\mbox{.}(2020)]%
        {Peng2020ConvolutionalON}
\bibfield{author}{\bibinfo{person}{Songyou Peng}, \bibinfo{person}{Michael Niemeyer}, \bibinfo{person}{Lars~M. Mescheder}, \bibinfo{person}{Marc Pollefeys}, {and} \bibinfo{person}{Andreas Geiger}.} \bibinfo{year}{2020}\natexlab{}.
\newblock \showarticletitle{Convolutional Occupancy Networks}.
\newblock \bibinfo{journal}{\emph{ArXiv}}  \bibinfo{volume}{abs/2003.04618} (\bibinfo{year}{2020}).
\newblock
\urldef\tempurl%
\url{https://api.semanticscholar.org/CorpusID:212646575}
\showURL{%
\tempurl}


\bibitem[Sang et~al\mbox{.}(2025)]%
        {Sang2025_Normals}
\bibfield{author}{\bibinfo{person}{Lu Sang}, \bibinfo{person}{Abhishek Saroha}, \bibinfo{person}{Maolin Gao}, {and} \bibinfo{person}{Daniel Cremers}.} \bibinfo{year}{2025}\natexlab{}.
\newblock \showarticletitle{Enhancing Surface Neural Implicits with Curvature-Guided Sampling and Uncertainty-Augmented Representations}. In \bibinfo{booktitle}{\emph{Pattern Recognition: 46th DAGM German Conference, DAGM GCPR 2024, Munich, Germany, September 10–13, 2024, Proceedings, Part I}} (Munich, Germany). \bibinfo{publisher}{Springer-Verlag}, \bibinfo{address}{Berlin, Heidelberg}, \bibinfo{pages}{312–328}.
\newblock
\showISBNx{978-3-031-85180-3}
\urldef\tempurl%
\url{https://doi.org/10.1007/978-3-031-85181-0_20}
\showDOI{\tempurl}


\bibitem[Selvaraju(2024)]%
        {Selvaraju2024_Developa}
\bibfield{author}{\bibinfo{person}{Pratheba Selvaraju}.} \bibinfo{year}{2024}\natexlab{}.
\newblock \showarticletitle{Developability Approximation for Neural Implicits Through Rank Minimization}. In \bibinfo{booktitle}{\emph{2024 International Conference on 3D Vision (3DV)}}. \bibinfo{pages}{780--789}.
\newblock
\urldef\tempurl%
\url{https://doi.org/10.1109/3DV62453.2024.00041}
\showDOI{\tempurl}


\bibitem[Sharma et~al\mbox{.}(2017)]%
        {Sharma2017CSGNetNS}
\bibfield{author}{\bibinfo{person}{Gopal Sharma}, \bibinfo{person}{Rishabh Goyal}, \bibinfo{person}{Difan Liu}, \bibinfo{person}{Evangelos Kalogerakis}, {and} \bibinfo{person}{Subhransu Maji}.} \bibinfo{year}{2017}\natexlab{}.
\newblock \showarticletitle{CSGNet: Neural Shape Parser for Constructive Solid Geometry}.
\newblock \bibinfo{journal}{\emph{2018 IEEE/CVF Conference on Computer Vision and Pattern Recognition}} (\bibinfo{year}{2017}), \bibinfo{pages}{5515--5523}.
\newblock
\urldef\tempurl%
\url{https://api.semanticscholar.org/CorpusID:4939249}
\showURL{%
\tempurl}


\bibitem[Sharma et~al\mbox{.}(2020)]%
        {Sharma2020ParSeNetAP}
\bibfield{author}{\bibinfo{person}{Gopal Sharma}, \bibinfo{person}{Difan Liu}, \bibinfo{person}{Evangelos Kalogerakis}, \bibinfo{person}{Subhransu Maji}, \bibinfo{person}{Siddhartha Chaudhuri}, {and} \bibinfo{person}{Radom'ir Mvech}.} \bibinfo{year}{2020}\natexlab{}.
\newblock \showarticletitle{ParSeNet: A Parametric Surface Fitting Network for 3D Point Clouds}.
\newblock \bibinfo{journal}{\emph{ArXiv}}  \bibinfo{volume}{abs/2003.12181} (\bibinfo{year}{2020}).
\newblock
\urldef\tempurl%
\url{https://api.semanticscholar.org/CorpusID:214692999}
\showURL{%
\tempurl}


\bibitem[Sitzmann et~al\mbox{.}(2020)]%
        {Sitzmann2020ImplicitNR}
\bibfield{author}{\bibinfo{person}{Vincent Sitzmann}, \bibinfo{person}{Julien N.~P. Martel}, \bibinfo{person}{Alexander~W. Bergman}, \bibinfo{person}{David~B. Lindell}, {and} \bibinfo{person}{Gordon Wetzstein}.} \bibinfo{year}{2020}\natexlab{}.
\newblock \showarticletitle{Implicit Neural Representations with Periodic Activation Functions}.
\newblock \bibinfo{journal}{\emph{ArXiv}}  \bibinfo{volume}{abs/2006.09661} (\bibinfo{year}{2020}).
\newblock
\urldef\tempurl%
\url{https://api.semanticscholar.org/CorpusID:219720931}
\showURL{%
\tempurl}


\bibitem[Tancik et~al\mbox{.}(2020)]%
        {Tancik2020FourierFL}
\bibfield{author}{\bibinfo{person}{Matthew Tancik}, \bibinfo{person}{Pratul~P. Srinivasan}, \bibinfo{person}{Ben Mildenhall}, \bibinfo{person}{Sara Fridovich-Keil}, \bibinfo{person}{Nithin Raghavan}, \bibinfo{person}{Utkarsh Singhal}, \bibinfo{person}{Ravi Ramamoorthi}, \bibinfo{person}{Jonathan~T. Barron}, {and} \bibinfo{person}{Ren Ng}.} \bibinfo{year}{2020}\natexlab{}.
\newblock \showarticletitle{Fourier Features Let Networks Learn High Frequency Functions in Low Dimensional Domains}.
\newblock \bibinfo{journal}{\emph{ArXiv}}  \bibinfo{volume}{abs/2006.10739} (\bibinfo{year}{2020}).
\newblock
\urldef\tempurl%
\url{https://api.semanticscholar.org/CorpusID:219791950}
\showURL{%
\tempurl}


\bibitem[Tang et~al\mbox{.}(2021)]%
        {Tang2021OctFieldHI}
\bibfield{author}{\bibinfo{person}{Jiapeng Tang}, \bibinfo{person}{Weikai Chen}, \bibinfo{person}{Jie Yang}, \bibinfo{person}{Bo Wang}, \bibinfo{person}{Songrun Liu}, \bibinfo{person}{Bo Yang}, {and} \bibinfo{person}{Lin Gao}.} \bibinfo{year}{2021}\natexlab{}.
\newblock \showarticletitle{OctField: Hierarchical Implicit Functions for 3D Modeling}. In \bibinfo{booktitle}{\emph{Neural Information Processing Systems}}.
\newblock
\urldef\tempurl%
\url{https://api.semanticscholar.org/CorpusID:240354073}
\showURL{%
\tempurl}


\bibitem[Uy et~al\mbox{.}(2021)]%
        {Uy2021Point2CylRE}
\bibfield{author}{\bibinfo{person}{Mikaela~Angelina Uy}, \bibinfo{person}{Yen-Yu Chang}, \bibinfo{person}{Minhyuk Sung}, \bibinfo{person}{Purvi Goel}, \bibinfo{person}{J. Lambourne}, \bibinfo{person}{Tolga Birdal}, {and} \bibinfo{person}{Leonidas~J. Guibas}.} \bibinfo{year}{2021}\natexlab{}.
\newblock \showarticletitle{Point2Cyl: Reverse Engineering 3D Objects from Point Clouds to Extrusion Cylinders}.
\newblock \bibinfo{journal}{\emph{2022 IEEE/CVF Conference on Computer Vision and Pattern Recognition (CVPR)}} (\bibinfo{year}{2021}), \bibinfo{pages}{11840--11850}.
\newblock
\urldef\tempurl%
\url{https://api.semanticscholar.org/CorpusID:245329808}
\showURL{%
\tempurl}


\bibitem[Wang et~al\mbox{.}(2022)]%
        {Wang2022DualOG}
\bibfield{author}{\bibinfo{person}{Peng-Shuai Wang}, \bibinfo{person}{Yang Liu}, {and} \bibinfo{person}{Xin Tong}.} \bibinfo{year}{2022}\natexlab{}.
\newblock \showarticletitle{Dual octree graph networks for learning adaptive volumetric shape representations}.
\newblock \bibinfo{journal}{\emph{ACM Transactions on Graphics (TOG)}}  \bibinfo{volume}{41} (\bibinfo{year}{2022}), \bibinfo{pages}{1 -- 15}.
\newblock
\urldef\tempurl%
\url{https://api.semanticscholar.org/CorpusID:248525003}
\showURL{%
\tempurl}


\bibitem[Wang et~al\mbox{.}(2023)]%
        {Wang2023NeuralSingularHessianIN}
\bibfield{author}{\bibinfo{person}{Zixiong Wang}, \bibinfo{person}{Yunxiao Zhang}, \bibinfo{person}{Rui Xu}, \bibinfo{person}{Fan Zhang}, \bibinfo{person}{Peng Wang}, \bibinfo{person}{Shuangmin Chen}, \bibinfo{person}{Shiqing Xin}, \bibinfo{person}{Wenping Wang}, {and} \bibinfo{person}{Changhe Tu}.} \bibinfo{year}{2023}\natexlab{}.
\newblock \showarticletitle{Neural-Singular-Hessian: Implicit Neural Representation of Unoriented Point Clouds by Enforcing Singular Hessian}.
\newblock \bibinfo{journal}{\emph{ACM Transactions on Graphics (TOG)}}  \bibinfo{volume}{42} (\bibinfo{year}{2023}), \bibinfo{pages}{1 -- 14}.
\newblock
\urldef\tempurl%
\url{https://api.semanticscholar.org/CorpusID:261529950}
\showURL{%
\tempurl}


\bibitem[Ye et~al\mbox{.}(2022)]%
        {Ye2022GIFSNI}
\bibfield{author}{\bibinfo{person}{Jianglong Ye}, \bibinfo{person}{Yuntao Chen}, \bibinfo{person}{Naiyan Wang}, {and} \bibinfo{person}{X. Wang}.} \bibinfo{year}{2022}\natexlab{}.
\newblock \showarticletitle{GIFS: Neural Implicit Function for General Shape Representation}.
\newblock \bibinfo{journal}{\emph{2022 IEEE/CVF Conference on Computer Vision and Pattern Recognition (CVPR)}} (\bibinfo{year}{2022}), \bibinfo{pages}{12819--12829}.
\newblock
\urldef\tempurl%
\url{https://api.semanticscholar.org/CorpusID:248177781}
\showURL{%
\tempurl}


\end{thebibliography}
    \balance
}
{
}
	
\end{document}